\documentclass[,useAMS,usenatbib]{mn2e}

\usepackage{graphicx}

\title[FeLoBAL outflow constraints from absorption line variability]{Constraining FeLoBAL outflows from absorption line variability}
\author[S. M. McGraw et al.]{S. M. McGraw,$^{1}$ J. C. Shields,$^{1}$ F. W. Hamann,$^{2}$ D. M. Capellupo,$^{3}$ S. C. Gallagher,$^{4,5}$ 
\newauthor
and W. N. Brandt$^{6,7,8}$ \\
$^{1}$Department of Physics and Astronomy, Ohio University, Athens, Ohio 45701, USA \\
$^{2}$Department of Astronomy, University of Florida, Gainesville, Florida 32611, USA \\
$^{3}$Department of Astrophysics, Tel Aviv University, Tel Aviv 69978, Israel \\
$^{4}$Department of Physics and Astronomy, University of Western Ontario, London, Ontario N6A 3K7, Canada \\
$^{5}$Yale Center for Astronomy and Astrophysics, Department of Physics, Yale University, New Haven, Connecticut 06520, USA \\
$^{6}$Department of Astronomy \& Astrophysics, The Pennsylvania State University, University Park, Pennsylvania 16802, USA \\
$^{7}$Institute for Gravitation and the Cosmos, The Pennsylvania State University, University Park, Pennsylvania 16802, USA \\
$^{8}$Department of Physics, The Pennsylvania State University, University Park, Pennsylvania 16802, USA}
\begin{document}

\date{\today}

\pagerange{\pageref{firstpage}--\pageref{lastpage}} \pubyear{2002}

\maketitle

\label{firstpage}

\begin{abstract}

FeLoBALs are a rare class of quasar outflows with low-ionization broad absorption lines (BALs), large column densities, and potentially large kinetic energies that might be important for `feedback' to galaxy evolution. In order to probe the physical properties of these outflows, we conducted a multiple-epoch, absorption line variability study of 12 FeLoBAL quasars spanning a redshift range of $0.7 \le z \le 1.9$ over rest frame time-scales of $\sim$10 d to 7.6 yr. We detect absorption line variability with $\ge$ 8$\sigma$ confidence in 3 out of the 12 sources in our sample over time-scales of $\sim$0.6 to 7.6 yr. Variable wavelength intervals are associated with ground and excited state Fe\thinspace \textsc{ii} multiplets, the Mg\thinspace \textsc{ii} $\lambda\lambda$2796, 2803 doublet, Mg\thinspace \textsc{i} $\lambda$2852, and excited state Ni\thinspace \textsc{ii} multiplets. The observed variability along with evidence of saturation in the absorption lines favors transverse motions of gas across the line of sight (LOS) as the preferred scenario, and allows us to constrain the outflow distance from the supermassive black hole (SMBH) to be $\la69$, 7, and 60 pc for our three variable sources. In combination with other studies, these results suggest that the outflowing gas in FeLoBAL quasars resides on a range of scales and includes matter within tens of parsecs of the central source.

\end{abstract}

\begin{keywords}
galaxies: active -- quasars: general -- quasars: absorption lines.
\end{keywords}

\section{Introduction}

Broad absorption lines (BALs) in rest frame UV quasar spectra are typically seen as blue-shifted troughs that indicate the existence of gas outflows. The ions that produce the absorption are often highly ionized species such as C\thinspace \textsc{iv}, N\thinspace \textsc{v}, and O\thinspace \textsc{vi}; their presence defines the high-ionization BAL (HiBAL) quasar class, which comprises between $\sim$10 and 26 per cent of quasars \citep{tru06}. Sources with additional BALs from lower ionized species (e.g. Mg\thinspace \textsc{ii}, Al\thinspace \textsc{ii}, and Al\thinspace \textsc{iii}) are categorized as low-ionization BALs (LoBALs) and compose a much smaller fraction of quasars than HiBALs.

Throughout this paper, we define iron low-ionization BALs, or FeLoBALs, as any quasars with Fe\thinspace \textsc{ii} absorption along with transitions that characterize HiBALs and LoBALs (see \citealt{hal02a,hal02b} and references therein for general FeLoBAL information). \citet{tru06} estimated FeLoBALs to comprise $\la0.3$ per cent\footnote{The absorbing gas and dust likely produce selection effects in a flux-limited sample, and 0.3 per cent may therefore not represent the true FeLoBAL quasar fraction.} of quasars. The rarity of FeLoBAL outflows is not well understood; they may represent a short-lived phase of quasar evolution, or perhaps the Fe\thinspace \textsc{ii}-Fe\thinspace \textsc{iii} gas covers a very small fraction of the active galactic nucleus (AGN) and only few sight lines permit their detection \citep{ham04,ham12}.

There are conflicting ideas about where FeLoBAL outflows are located, and this has important implications for the calculated kinetic energies and the possible role of these outflows in `feedback' processes with the host galaxy. FeLoBALs at kiloparsec distances from the supermassive black hole (SMBH) may inject enough energy and momentum into the interstellar medium (ISM) to disrupt star formation (\citealt{far12, lei14}). \citet{fau12a} and \citet{fau12b} provide a theoretical framework for the formation of Fe\thinspace \textsc{ii} and Fe\thinspace \textsc{iii} at such radial scales that consists of gas driven from the accretion disk interacting with ISM clumps. Other models (\citealt{mur95,dek95}) instead suggest that all ionic species arise from gas accelerated off the accretion disk at sub-parsec scales from the SMBH. 

Studies that have utilized unblended, excited state Fe\thinspace \textsc{ii} lines and photoionization models have constrained FeLoBAL outflows to exist $\ga700$ pc from the SMBH in a handful of sources, with estimated kinetic luminosities and mass flow rates on the order of  $\sim$1 per cent of $L_{\rmn{bol}}$ and $\sim$100 M$_{\sun}$ yr$^{-1}$, respectively (\citealt{dek01,kor08,moe09,dun10,bau10}). Kinetic luminosities and mass flow rates of these magnitudes make these sources  viable contributors to feedback and may play a role in explaining SMBH--bulge relations (e.g. $M$--$\sigma$ relation, \citealt{fer00,geb00}).  

Observational constraints on outflow location and other physical quantities can also be extracted using absorption line variability studies. Two scenarios that can potentially explain variations are that the absorbing gas undergoes an ionization change, or there is a change in covering fraction along our line of sight (LOS) due to transverse motions of gas across the emission region of the AGN (see \S 5.1 of \citealt{cap13} for other possibilities and reasons why they are unlikely for most sources). Limits on outflow location depend on which interpretation is utilized. An ionization change relates the measured variability time-scale to the recombination time of the gas, allowing for an estimate of the density, and hence distance from the central source. Models invoking absorbers crossing the LOS constrain an outflow crossing speed from the measured time-scale, which can be related to a keplerian velocity and hence distance from the central black hole.

Work focused on BAL variability associated with high-ionization species has established a range of outflow distances. \citet{moe09} and \citet{cap13} used models involving absorbers crossing the LOS to constrain the outflows to be at most sub-parsec and parsec scales from the SMBH, respectively, while \citet{bar94} assumed an ionization change scenario and established an upper limit of a few hundred parsecs. \citet{cap13} discussed the relative merits of the two interpretations, and \citet{cap14a} provided strong evidence supporting a LOS--crossing scenario in the P\thinspace \textsc{v} BAL quasar Q1413+1143.

Very few published studies of FeLoBAL variability exist. The properties of the entire FeLoBAL population are therefore not well understood due to few constraints on the outflows and a lack of statistical results from large sample variability studies. \citet{hal11} observed dramatic changes in Mg\thinspace \textsc{ii} and a disappearance of the Fe\thinspace \textsc{ii} lines at the same velocity in FBQS J1408+3054, and constrained the outflow to be between 1.7 and 14 pc from the SMBH based upon a model with clouds moving across the LOS. \citet{viv12} probed the time variability of five FeLoBAL sources, reporting changes from Fe\thinspace \textsc{iii} and Al\thinspace \textsc{iii} fine structure absorption lines in one of those objects and concluded that those ions likely originated in the ISM based upon ionization considerations and a model from \citet{rog11}.

We have acquired multiple-epoch spectroscopy for a sample of twelve FeLoBAL quasars and report results on absorption line variability using objective statistical criteria. Due to the limited work on FeLoBAL variability to date, this study provides a significant increase in the information available on these outflows. Section 2 (\S 2) details the observations and data reduction procedures, \S 3 explains the analytical approach used for investigating variability, \S 4 presents our results, \S 5 discusses the implications of the variability, \S 6 summarizes our findings, and \S A provides details on individual objects of interest. 

\section{Data}

Our FeLoBAL quasar sample represents a subset of objects taken from catalogs compiled by \citet{tru06} and \citet{gib09} along with J121442\footnote{We will identify sources using the first part of their full names, listed in Table 1.}, previously studied by \citet{dek02a}, and J142703 (see \citealt{bec00}). \citet{tru06} identified FeLoBALs by visually inspecting their sample of BAL quasars in a redshift range $0.5 \le z \le 4.38$, based on the Sloan Digital Sky Survey (SDSS) data release 3. \citet{gib09} constructed a sample of LoBALs with redshift  $0.48 \le z \le 2.28$ from SDSS data release 5, and we further identified FeLoBALs from visual inspection of those sources. The selection process involved identifying quasars with strong, low-ionization Fe\thinspace \textsc{ii} or Fe\thinspace \textsc{iii} absorption lines and proceeded without any prior knowledge of whether a given source was variable. Due to selection effects, the conclusions of this study may not apply to the entire FeLoBAL population.

\begin{table*}
 \centering
 \begin{minipage}{210mm}
  \caption{FeLoBAL quasar sample.}
  \begin{tabular}{@{}cccccccccc@{}}
  \hline
 Name & $z$\footnote{Emission line redshifts taken from \citet{hew10}.} & $M_{i}$\footnote{Absolute $i-$band magnitude corrected to $z=2$ (\citealt{she11}).} & HET\footnote{Observations are in decimal years. Dates in parentheses indicate a weighted average of the years preceding it.} & DR7 & DR12 & MDM & Coverage (\AA)\footnote{Represents the maximum, shared wavelength coverage (rest frame) between two spectra in each source.} & Variable?\footnote{Indicates whether absorption line variability was detected with the 8$\sigma$ or 6$\sigma$ criterion [yes (Y), no (N), or tentative (T)].  See \S 3.2 for details.} & $R$\footnote{Radio loudness (i.e. $R \equiv f_{\rmn{6 cm}}/f_{\rmn{2500}}$) taken from \citet{she11}.} \\
  \hline
J030000.57+004828.0 & 0.9003 & -27.6 & --- & 2000.73, 2000.92,  & --- & 2012.07 & 2000-3130 & N & 0.0 \\
&&&& 2001.81, (2001.14) &&&& \\
J031856.62--060037.7 & 1.9274 & -28.5 & --- & 2001.04 & --- & 2011.02, 2012.08 & 1300-2030 & N & --- \\
J084044.41+363327.8 & 1.2253 & -28.4 & 2000.10, 2000.20, & 2002.13 & --- & 2010.21, 2011.03, & 1930--3260 & T & 4.6 \\
&&& 2000.28, 2000.34, &&& 2012.08 && \\
&&& (2000.23) &&&&&& \\
J104459.61+365605.2 & 0.7010 & -26.8 & --- & 2005.25 & 2011.15 & 2011.02, 2012.07 & 2230-5410 & T & 22 \\
J112220.77+153927.9 & 1.1083 & -26.1 & --- & 2005.03 & 2012.16 & 2012.09 & 1800-4360 & T & 0.0 \\
J112526.12+002901.3 & 0.8635 & -25.8 & --- & 2000.19 & 2011.04 & 2012.07 & 2040-4940 & T & 0.0 \\
J115436.60+030006.3 & 1.3904 & -27.1 & --- & 2001.38 & 2011.31 & 2012.08 & 1590-3850 & N & 0.0 \\
J120337.92+153006.7 & 1.2176 & -26.9 & --- & 2005.27 & --- & 2012.09 & 1800-3090 & Y & 5.4 \\
J121442.30+280329.0 & 0.6949 & -26.3 & 2000.17, 2000.24, & 2006.24 & 2013.18 & 2010.21, 2011.31, & 2540-4280 & Y & 14 \\
&&& 2000.33, 2000.46,  &&& 2012.08 && \\
&&& (2000.28) &&&&&& \\
J123103.71+392903.7 & 1.0038 & -25.7 & --- & 2005.26 & 2011.08 & 2012.08, 2013.05 & 1900-4590 & Y & 0.0 \\
J142703.60+270940.0 & 1.1661 & -26.3 & 2000.17, 2000.24, & 2006.38 & 2012.38 & 2010.14, 2010.36, & 1750-4250 & T & 24 \\
&&& 2000.33, 2000.53, &&& 2011.31 && \\
&&& (2000.26) &&&&&& \\
J152438.80+415543.0 & 1.2301 &-26.3  & --- & 2005.39 & 2012.17 & 2012.09 & 1700-4130 & N & 22 \\
\hline
\end{tabular}
\end{minipage}
\end{table*} 

The twelve FeLoBAL quasars in our sample are listed in Table 1 with relevant information including redshift, absolute $i$-band magnitude, dates of observation, wavelength coverage, variability, and radio loudness. The range of redshift across our sample is $0.7 \le z \le 1.9$ with time-scales between any two observations ranging from $\sim$10 d to 7.6 yr in the quasar's rest frame\footnote{All subsequent time-scales are in the rest frame of the quasar.}. Spectra were taken by the 2.4-m Hiltner telescope at MDM Observatory and the 9.2-m Hobby--Eberly Telescope (HET) at McDonald Observatory, and were supplemented by spectra from SDSS data release 7 (DR7) and 12 (DR12). We also utilized $V$-band magnitudes, estimated on the basis of an unfiltered CCD, from the Catalina Sky Survey (CSS; \citealt{dra09}) data release 2 in our interpretations of the variability (see \S 5.1.1).

MDM spectra from 2010 and 2013 were obtained using the Boller \& Chivens CCD Spectrograph (CCDS)\footnote{http://www.astronomy.ohio-state.edu/MDM/CCDS/} while MDM epochs from 2011--2012 employed the Ohio State Multi-Object Spectrograph (OSMOS; \citealt{mar11}). Slit widths used during observations with CCDS and OSMOS were 1 and 1.2 arcsec, respectively, and the slit angle was rotated to coincide with the parallactic angle to minimize effects of atmospheric dispersion. Integrated exposure times ranged between 2 and 3.5 h for each source in the sample.

Observations taken at MDM observatory were reduced using standard routines within the image reduction and analysis facility (\textsc{iraf})\footnote{\textsc{iraf} is distributed by the National Optical Astronomy Observatory, which is operated by the Association of Universities for Research in Astronomy (AURA) under cooperative agreement with the National Science Foundation.}. Processing steps included subtracting off bias structure, flat fielding the data to remove pixel-to-pixel variations in sensitivity, extraction of spectra and their associated error arrays\footnote{Extracted error spectra account for CCD read noise and photon counting statistics.}, binning the dispersion axis into wavelength units, and applying an extinction correction and flux calibration to the spectra using a standard star from the same night. While the relative flux calibration of our spectra shows good reproducibility, the absolute flux calibration is less reliable in some cases due to variable observing conditions. Multiple exposures of a given object were combined using an average weighted by the inverse of the normalized variance of each exposure.

Spectra from HET were acquired using the low-resolution spectrograph (LRS; \citealp{hil98}). Observations were carried out at an average airmass of $\sim$1.2, with exposure times ranging from 660--1920 s. The data were processed using standard reduction procedures and were not flux calibrated. 

\begin{table}
\caption{Spectral parameters.}
\begin{tabular}{@{}cccc}
\hline
Instrument & $\lambda$-coverage (\AA) & Pixel width (km s$^{-1}$) & Resolution (km s$^{-1})$ \\
\hline
DR7 & 3800--9200 & 70 & 150 \\
DR12 & 3600--10400 & 70 & 140 \\
OSMOS & 3100--6800 & 40 & 190 \\
CCDS & ... & 80 & 200 \\
LRS & 4250--7400 & 50 & 230 \\
\hline
\end{tabular}
\medskip

Pixel widths and FWHM resolutions represent averages over the measured wavelength intervals. Also note that CCDS observations spanned 1600 \AA \ in a single exposure and the coverage window varied between sources.
\end{table}  

To facilitate comparisons of narrow lines, spectra of a given source were smoothed to match the lowest resolution observation via convolution with a gaussian of appropriate width. Table 2 shows spectral resolutions representative of each instrument used to collect data for our sample. The wavelength calibrations were checked using the O\thinspace \textsc{i} $\lambda$5577 atmospheric line and spectra from our entire dataset were adjusted as necessary to achieve agreement. We also resampled all spectra to a standard width of 70 km s$^{-1}$ per pixel for our analysis due to varying dispersions between instruments (see Table 2).

\section{Analysis}

In order to study absorption line changes across two or more observations in a given source, our general procedure involves two steps:  
\begin{enumerate}
\renewcommand{\theenumi}{(\arabic{enumi})}
\item Scale non-DR7 spectra to match the DR7 spectrum at non-variable wavelengths using the ratios between the observations.
\item Quantify absorption line variability by taking flux differences between every pair of available spectra and comparing those differences with propagated errors.
\end{enumerate}

\subsection{Scaling spectra}

Spectra were scaled by first computing the ratio between DR7 and non-DR7 observations and inspecting the resulting ratio to determine which wavelength intervals exhibited no `features'. Features are defined here as abrupt (i.e. occurring over intervals $\la100$ \AA) changes in the gradient of the ratio, and are distinguishable from both the noise and global slope changes over a large portion of the wavelength coverage ($\ga1000$ \AA). Features in the ratio indicate possible absorption line variability and we therefore omitted these wavelength intervals when scaling the spectra.  

Changes in the gradient of the ratio occurring over $\ga1000$ \AA \ were considered featureless wavelength intervals and used to scale the spectra. They could be due to quasar continuum variations, systematic causes (i.e. flux calibration issues or loss of light due to atmospheric dispersion), or perhaps concurrent variability of overlapping BALs that may be confused with the two previous points. This last issue is not a problem for our sample for two reasons:   
\begin{enumerate}
\item In sources with overlapping BALs there are also wavelength regions free of absorption and a feature in the ratio would therefore be detectable at the onset of the BALs.
\item Spectra that suffer from absorption over the entire wavelength coverage exhibit BALs with structure and well-defined edges that would show up as features in the ratio.
\end{enumerate}

After identifying features in the ratio, fits were performed to the featureless intervals collectively over the entire wavelength coverage to produce a scaling function. Functions used in fitting ratios that included MDM and DR12 epochs ranged between constant values, first order\footnote{The order specifies the number of spline pieces used for the fit.} linear splines, and low order cubic splines; HET data called for cubic spline functions up to sixth order due to lack of flux calibration.    

When applicable, only featureless intervals not associated with absorption were chosen to serve as constraints for the fit to interpolate or extrapolate over absorption lines. There were exceptions where nearly the entire wavelength coverage was contaminated by absorption (J084044, J112220, J112526, J115436) or the lack of flux calibrated spectra from HET required fitting with a non-linear function (J084044, J121442). If, in these situations, wavelength intervals containing absorption were featureless as described above, we included these regions when fitting in order to scale the spectra in a reasonable way. Two sources exhibited features across nearly the entire wavelength coverage and required a separate approach for analyzing the variability (J120337, J123103). 

The final step involved multiplying the non-DR7 spectra by the functions produced from fitting the ratios, which resulted in scaled non-DR7 spectra that coincided with the DR7 spectrum at wavelength intervals that were featureless in the ratios.  

\subsection{Quantifying absorption line variability}

After scaling all available spectra, we adopted a quantitative criterion to gauge whether variability was statistically significant when comparing two spectra from a given object. The significance of variability was calculated by taking the difference in flux between two observations, binning the flux difference spectrum by 280 km s$^{-1}$, and comparing each binned pixel with the propagated error $\sigma$ in that pixel. The 280 km s$^{-1}$ bin width (corresponding to 4 pixels sampled to 70 km s$^{-1}$) is greater than the lowest resolution in our dataset (see Table 2) and is meant to detect variations across a range of pixels instead of probing pixel-to-pixel fluctuations that may not reflect true variability.

Our quantitative criterion for significant variability was that the flux difference in a given binned pixel had to be at least eight times the propagated error value ($8\sigma$), either positive or negative. The propagated error ($\sigma$) only accounts for CCD noise and random photon statistics. The $8\sigma$ threshold accounts for the systematic uncertainties associated with the scaling process and correcting for resolution differences between spectra, and was determined based upon tests employing alternative fits to ratios along with convolving spectra with a gaussian of varying width.

Utilizing the  conservative criterion outlined above as the sole indicator of variability is questionable given the difficulty in quantifying the systematic uncertainties, particularly in scaling spectra, due to structural differences between the ratios.  In light of this we established a threshold for `tentative' variability to be six times the propagated error (6$\sigma$) for each binned pixel and proceeded to examine each source individually to place it into one of three categories:  Non-variable (N), tentative (T),  or variable (Y). Each source's category is listed in Table 1 and shown in Fig. 1. Each plot in Fig. 1 represents one of the sources in our sample with two spectra from different epochs plotted on top of one another for comparison. The pair of spectra from each source with the maximum, shared wavelength coverage over the absorption lines and the largest variability by eye is shown in Fig. 1.

\begin{figure*}
\begin{center}
\includegraphics[width=2.0\columnwidth,angle=0]{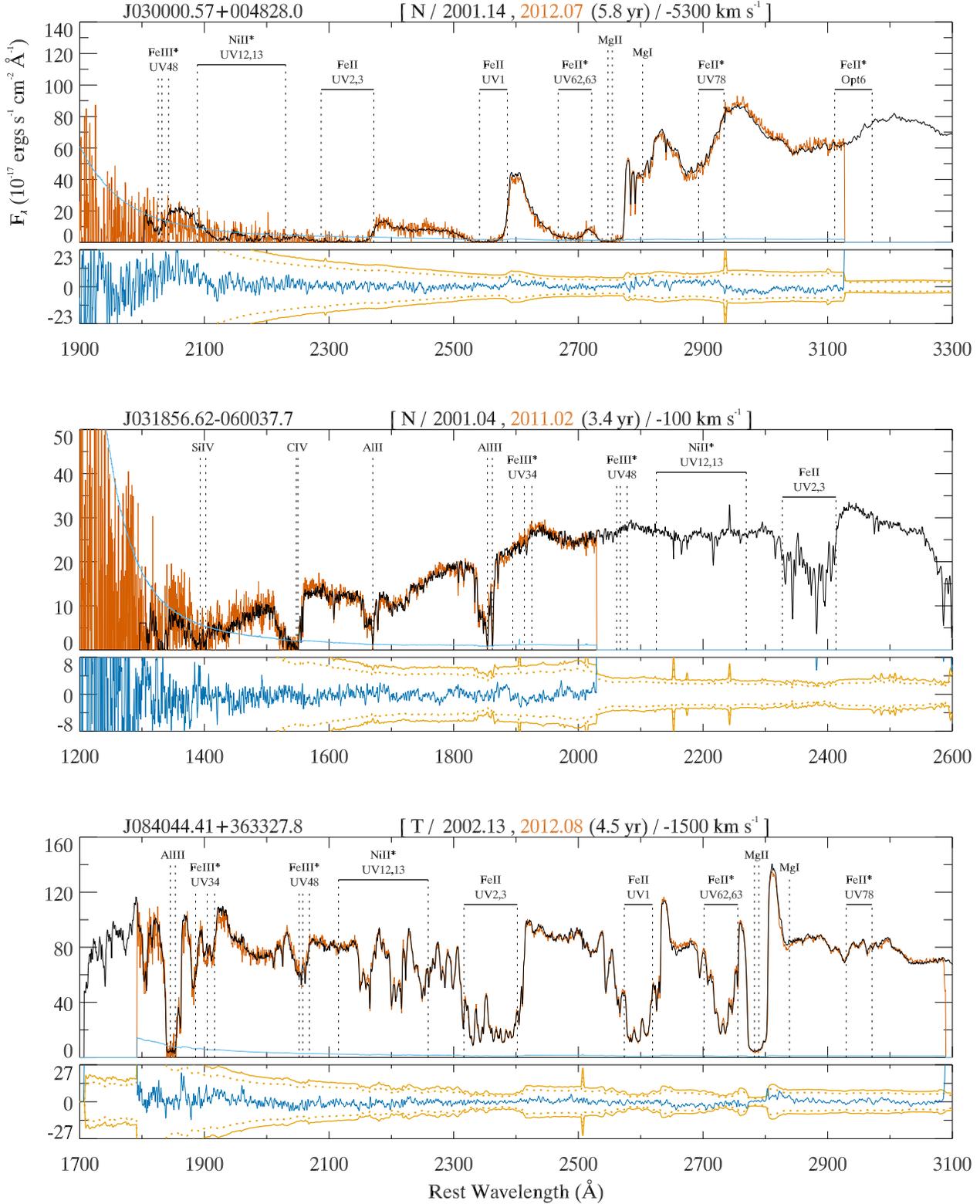}
\caption{FeLoBAL spectra comparisons. The top of each panel lists the object name, variability category, dates and color of the two epochs shown, the rest frame time-scale between the two epochs, and the velocity shift of the overlaid line template (negative velocity indicates the lines are blue-shifted). For the error arrays associated with the plotted spectra pairs, the one with larger amplitude is indicated by the light blue curve along the bottom of each large panel. The small panels show the binned difference spectrum (blue) along with the 8$\sigma$ (orange, solid) and 6$\sigma$ (orange, dashed) propagated noise spectra.  The order of the dates in each panel correspond to the order of the difference operation.}
\end{center}
\end{figure*}

\setcounter{figure}{1}

\begin{figure*}
\begin{center}
\includegraphics[width=2.0\columnwidth,angle=0]{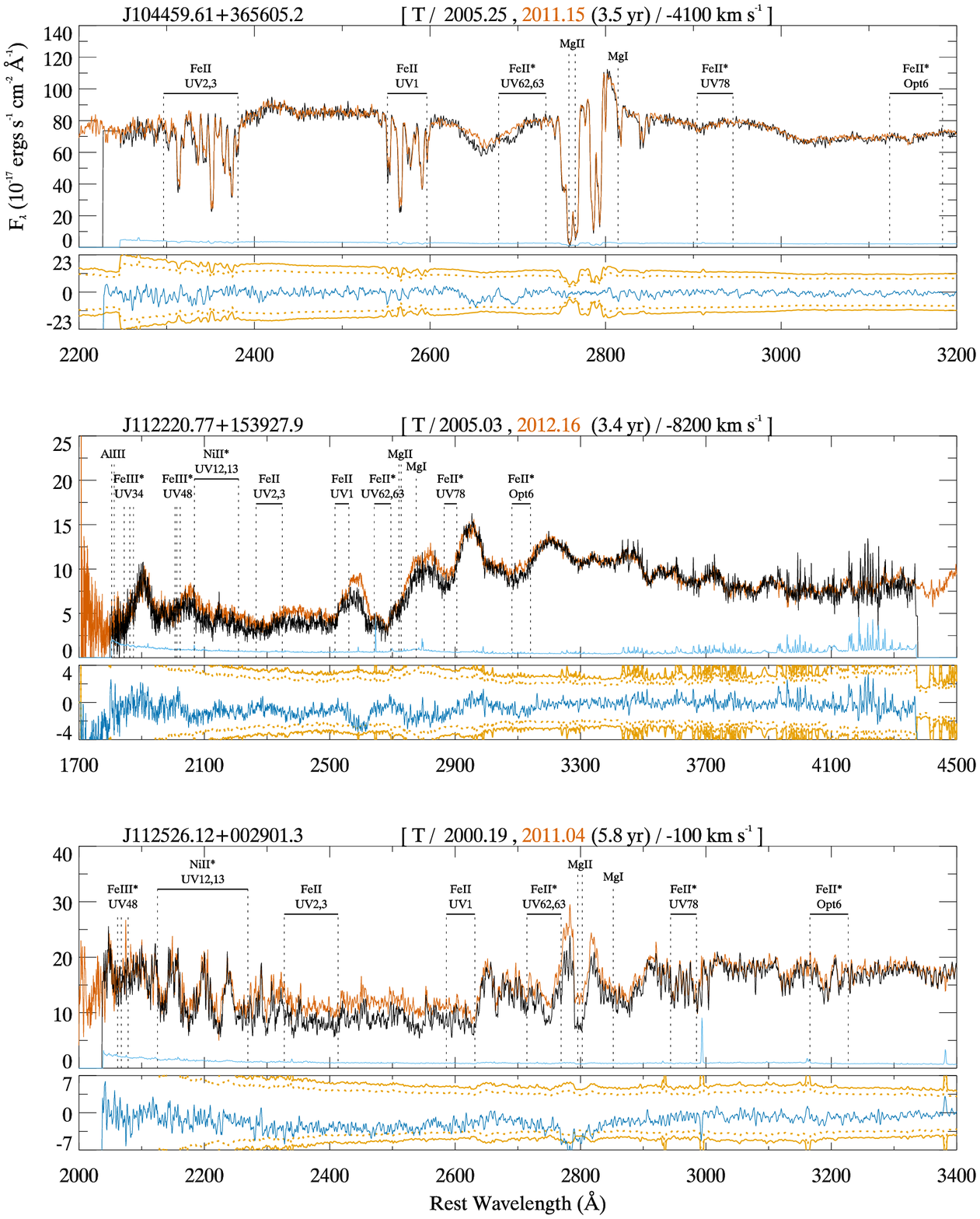}
\contcaption{}
\end{center}
\end{figure*}

\setcounter{figure}{1}

\begin{figure*}
\begin{center}
\includegraphics[width=2.0\columnwidth,angle=0]{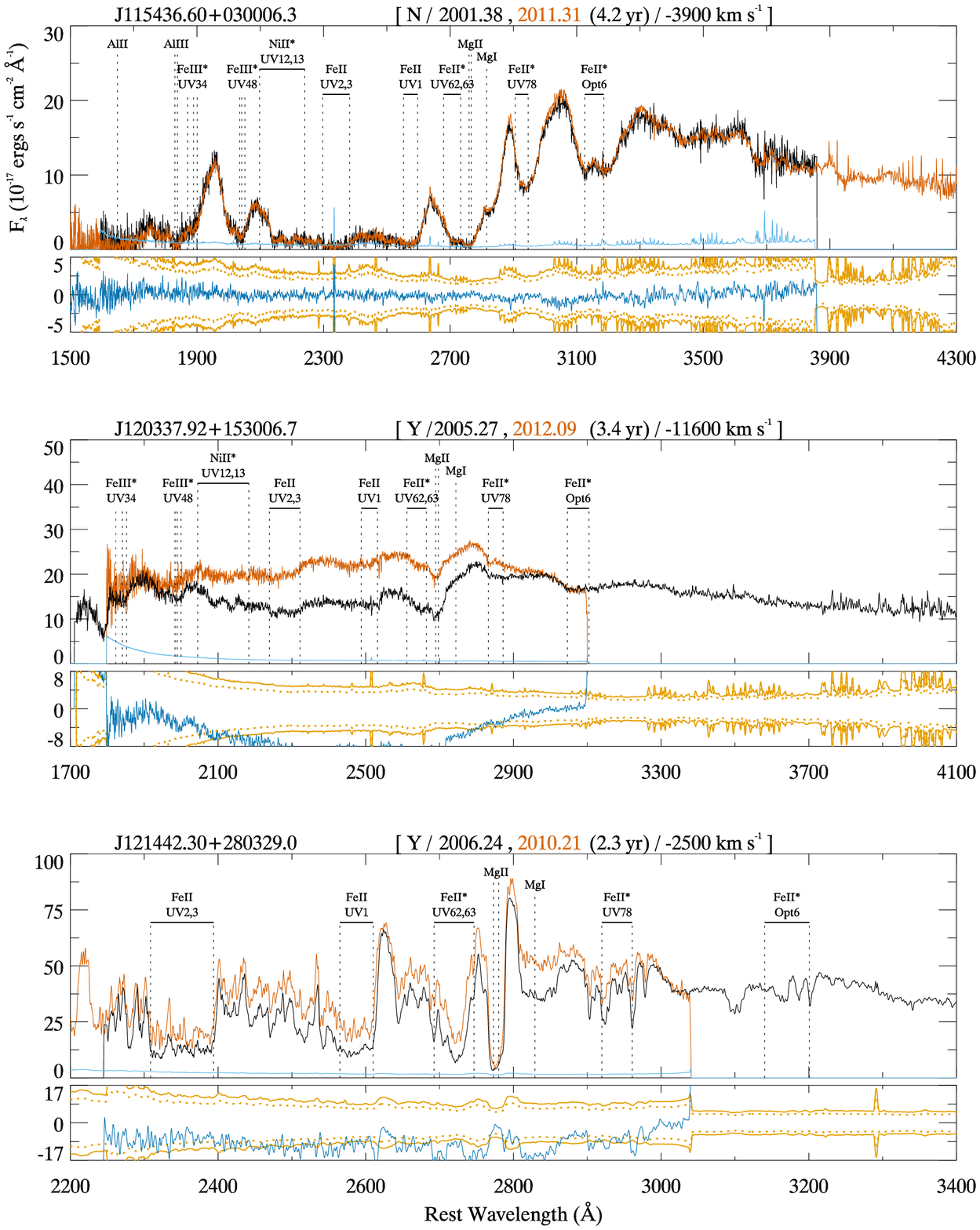}
\contcaption{}
\end{center}
\end{figure*}

\setcounter{figure}{1}

\begin{figure*}
\begin{center}
\includegraphics[width=2.0\columnwidth,angle=0]{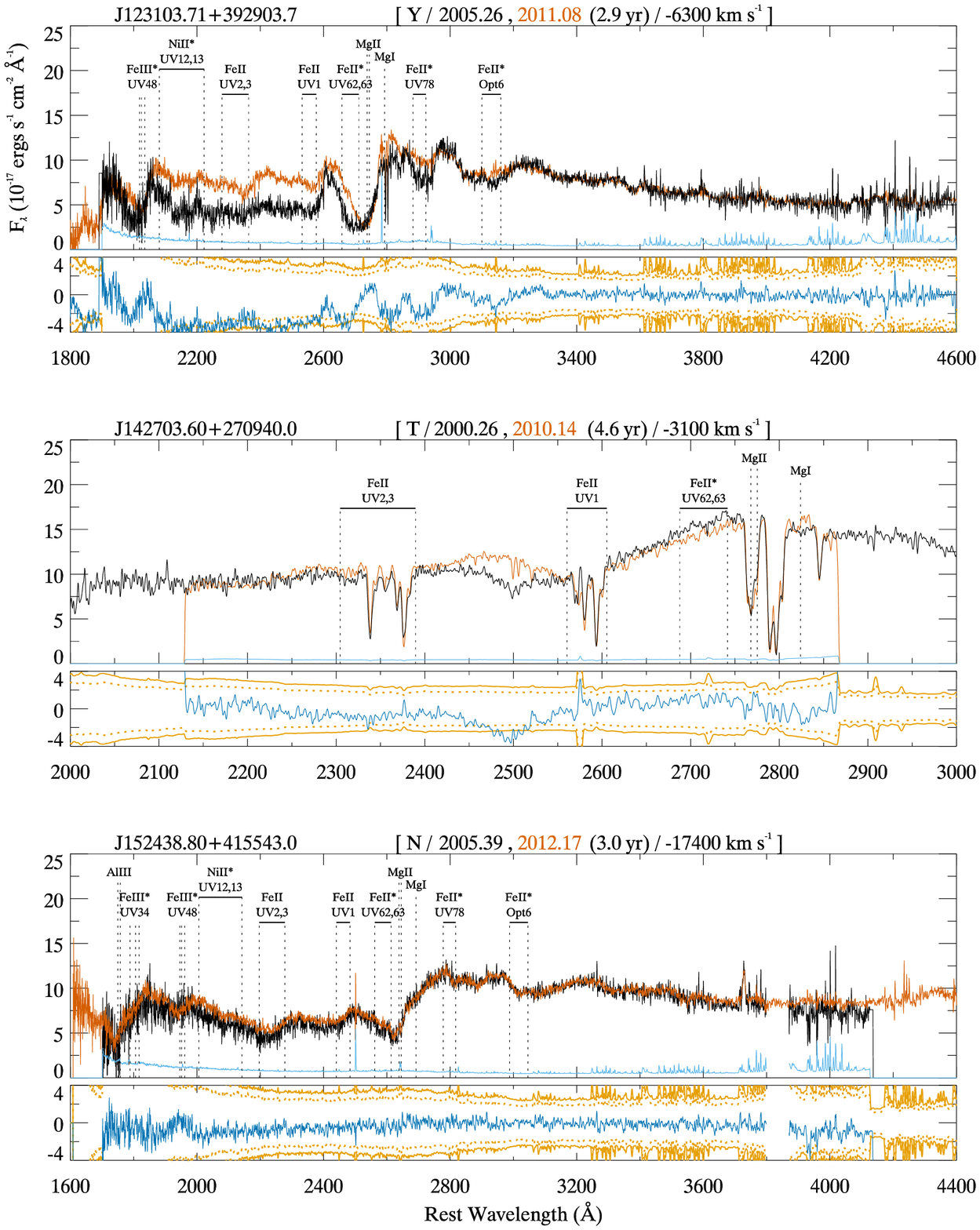}
\contcaption{}
\end{center}
\end{figure*}

Objects in category N did not meet either the $8\sigma$ or $6\sigma$ thresholds in any binned pixel. Comparisons that did meet either $8\sigma$ or $6\sigma$ yet further inspection by eye suggested the variations to be systematic were also conservatively placed in category N (see notes in \S A.3). Sources fell in the T category if they met the $6\sigma$ threshold in intervals associated with absorption with no apparent systematic issues (e.g. J112220), or met the $8\sigma$ threshold but the variability could not be attributed to absorption conclusively (e.g. J112526, J142703). Objects categorized as Y met the 8$\sigma$ criterion with none of the caveats listed for the N or T categories. The existence of underlying emission lines in AGN spectra (e.g. broad Fe\thinspace \textsc{ii} emission, \citealt{ves01}) can make some variable cases ambiguous, and we have taken this into consideration in our assessments of the variability.

For the purpose of identifying lines as an aid in interpreting absorption line variability, we generated a list of ionic transitions that is overlaid on all plots in each figure. The list was produced by inspecting all spectra in our sample and identifying transitions and multiplets associated with absorption, and is based on an inventory of transitions in BAL quasars compiled by \citet{hal02b}. We represent multiplets consisting of more than three transitions as horizontal bars extending from the lowest to highest transition wavelength. The list is not intended to identify all transitions contributing to the absorption in any source; line blending from the many transitions in Fe\thinspace \textsc{ii} and other ions make it difficult to identify these lines individually. The UV multiplet notation comes from \citet{moo50}.

\subsection{Associated absorption lines}

Two sources in our sample (J030000 and J123103) show detectable associated absorption lines (AALs) coming from Mg\thinspace \textsc{ii} $\lambda\lambda$2796, 2803 and Mg\thinspace \textsc{i} $\lambda$2852, as shown in Fig. 2 (see e.g. \citealt{she12} for recent work and general information on AALs). Following the procedure outlined above, variability between two observations was checked by scaling one spectrum to coincide with the other; for this analysis we only scaled using intervals around the Mg\thinspace \textsc{ii} and Mg\thinspace \textsc{i} AALs, and used either constants or first order linear splines as scaling functions.

We determined whether AAL variability was significant using similar statistical criteria as in the main analysis. The requirement for significant AAL variability was that the flux difference in a given binned pixel had to be at least six times the propagated error value (6$\sigma$), either positive or negative. Similarly, tentative variability required the binned flux difference to be at least 4$\sigma$. As discussed in \S 3.2, the propagated error $\sigma$ does not account for uncertainties in scaling the spectra or errors from smoothing spectra to correct for instrumental resolution differences. The 6$\sigma$ and 4$\sigma$ thresholds for AAL variability are lower than the 8$\sigma$ and 6$\sigma$ criteria used in the main analysis because there is less uncertainty in scaling spectra in the vicinity of narrow absorption lines that only span $\sim$10 \AA. The 6$\sigma$ threshold was estimated by performing tests involving multiple scaling functions, different intervals to constrain the fit, and smoothing spectra by various widths to quantify the systematic errors. The 4$\sigma$ threshold allows for tentative cases to be quantified objectively.

\begin{figure}
\begin{center}
\includegraphics[width=1.0\columnwidth,angle=0]{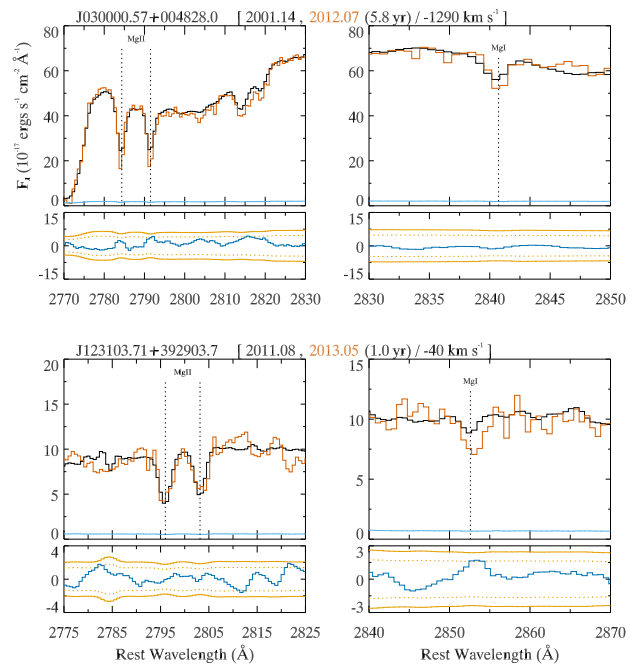}
\caption{AAL variability in J030000 and J123103.  Each source shows two panels, each depicting the Mg\thinspace \textsc{ii} $\lambda\lambda$2796, 2803 and Mg\thinspace \textsc{i} $\lambda$2852 AALs, respectively.  The only distinction compared to Fig. 1 is that the smaller panels show the 6$\sigma$ (orange, solid) and 4$\sigma$ (orange, dotted) error spectra plotted with the binned flux difference.}
\end{center}
\end{figure}

\section{Results}

\begin{figure*}
\begin{center}
\includegraphics[width=2.0\columnwidth,angle=0]{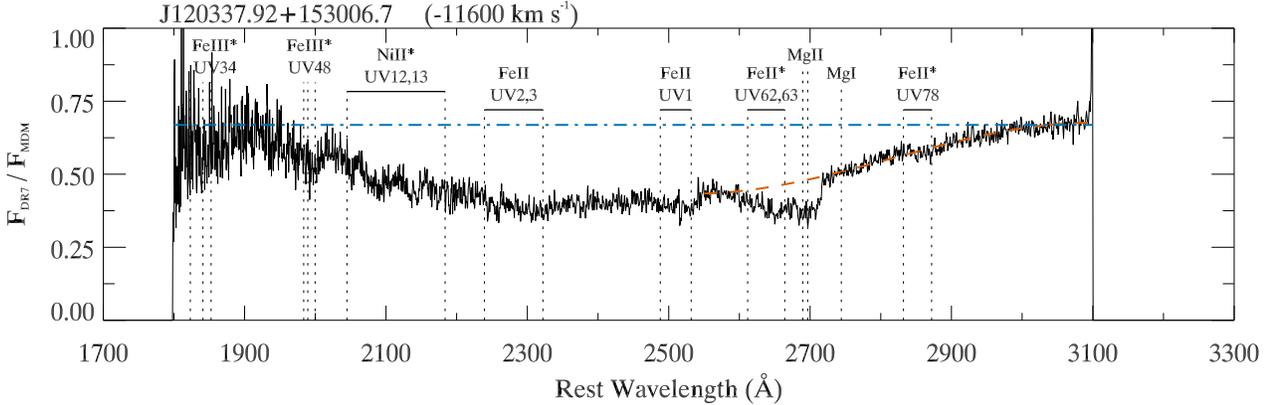}
\caption{Ratio between DR7 and MDM spectra for J120337, revealing evidence of variability in both continuum shape and absorption line strength.  The blue (dotted-dashed) and red (dashed) lines are the two functions that were used to scale the MDM epoch to coincide with DR7 (see \S 4.1), with the red curve optimized to fit longward of $\sim$2550 \AA.}
\end{center}
\end{figure*} 

We have detected significant broad absorption line variability from our 8$\sigma$ criterion in 3 out of the 12 FeLoBAL quasars in our sample and categorized them as Y sources (J120337, J121442, and J123103).  

The measured variability time-scales ranged from $\sim$0.6 to 7.6 yr. Absorption lines that appear to have varied come from Fe\thinspace \textsc{ii} (UV1,2,3), Fe\thinspace \textsc{ii}$^*$ (UV62,63,78)\footnote{An $^*$ after an ion signifies that the transition originated from an excited state.}, Mg\thinspace \textsc{ii} $\lambda\lambda 2796,2803$, Mg\thinspace \textsc{i} $\lambda$2852, and Ni\thinspace \textsc{ii}$^*$ (UV12,13) along with other Fe\thinspace \textsc{ii} multiplets and ionic species that are difficult to identify (see \S 4.2 for evidence of this). Other low-ionization lines that were identified but were not associated with regions that met the 8$\sigma$ criterion included Fe\thinspace \textsc{ii}$^*$ (Opt6), Fe\thinspace \textsc{iii}$^*$ (UV34,48), and Al\thinspace \textsc{iii} $\lambda\lambda1854, 1862$. One source in our sample (J031856) enabled us to examine BALs associated with C\thinspace \textsc{iv} $\lambda\lambda1548, 1550$, Si\thinspace \textsc{iv}$\lambda\lambda1393, 1402$ and Al\thinspace \textsc{ii} $\lambda$1671.

For the three Y sources we see evidence for absorption line variability across entire BAL troughs, portions of BAL troughs, and in narrower absorption lines that do not qualify as BALs. We calculated the average velocity width of variable regions meeting our 8$\sigma$ criterion within the Mg\thinspace \textsc{ii} $\lambda\lambda$ 2796, 2803 BAL to be $\sim$520 km s$^{-1}$; this average included all regions within the Mg\thinspace \textsc{ii} BAL, whether they overlapped with the adjacent Mg\thinspace \textsc{ii} broad emission line (BEL) or Fe\thinspace \textsc{ii}$^*$ (UV62,63) absorption.   

We also looked at how Mg\thinspace \textsc{ii} and Fe\thinspace \textsc{ii} varied in relation to one another. In all three of our Y objects we observed Mg\thinspace \textsc{ii} and Fe\thinspace \textsc{ii} BALs varying together, either increasing or decreasing in amplitude. J120337 exhibited decreases in absorption line strength over $\sim$3.1 yr across two observations; J123103 showed similar behavior over $\sim$2.9 yr. J121442 showed decreases in absorption strength followed by increases in absorption strength over six observations spanning $\sim$7.6 yr. The Y sources are discussed individually below. 

Table 1 also shows the radio loudness parameter $R$ (from \citealt{she11}, defined as the ratio of fluxes at 6 cm and 2500 \AA) for all available sources, and we observe no notable connection between radio loudness and degree of absorption line variability.

\subsection{J120337.92+153006.7}

Fig. 3 shows the ratio between the DR7 and MDM epochs in J120337. Inspection of Fig. 3 reveals indication of variability over nearly the entire wavelength coverage from $\sim$1940--3000 \AA. The global change is not a flux calibration issue because this source was observed on the same night as other objects that do not show any discrepancies in flux of this magnitude when compared to DR7. Slit losses from atmospheric dispersion are also likely unimportant given that this source was observed at very low airmass (i.e. $\la$ 1.1). Fluctuations due to quasar continuum emission are also unexpected as they typically occur over a much larger wavelength interval. Fig. 1 shows the resulting variability when the blue dot-dashed line in Fig. 3 is used to scale the MDM spectrum; variable regions meeting the 8$\sigma$ criterion range from $\sim$2090--2850 \AA.

It is plausible that the observed variations are due to highly blended absorption lines. Fig. 3 also shows localized, abrupt changes which coincide with and mimic the structure of the absorption lines in the DR7 spectrum in Fig. 1; the most notable changes in Fig. 3 occur at $\sim$2720, 2540, and 2010 \AA, corresponding to the Mg\thinspace \textsc{ii} $\lambda\lambda$2796, 2803, Fe\thinspace \textsc{ii} (UV1), and Fe\thinspace \textsc{iii}$^*$ (UV48) absorption lines, respectively. Lines that appear in the ratio between two spectra and exhibit the same structure as and align with the actual absorption lines imply that the absorption lines varied over the measured time-scales. However, the possibility remains that the observed variability from $\sim$1940--3000 \AA \  might also be due to peculiar continuum changes.

\begin{figure}
\begin{center}
\includegraphics[width=1.0\columnwidth,angle=0]{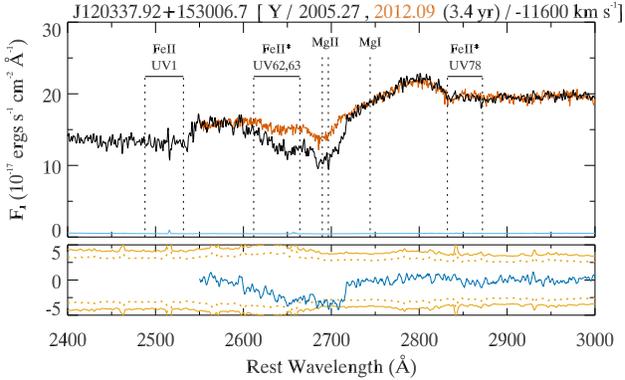}
\caption{Comparison between DR7 and rescaled MDM spectra for J120337 after removing a model of the large-scale variability, revealing changes  in the Mg\thinspace \textsc{ii} BAL.  See Fig. 1 for description of the panels.}
\end{center}
\end{figure} 

\begin{figure*}
\begin{center}
\includegraphics[width=2.0\columnwidth,angle=0]{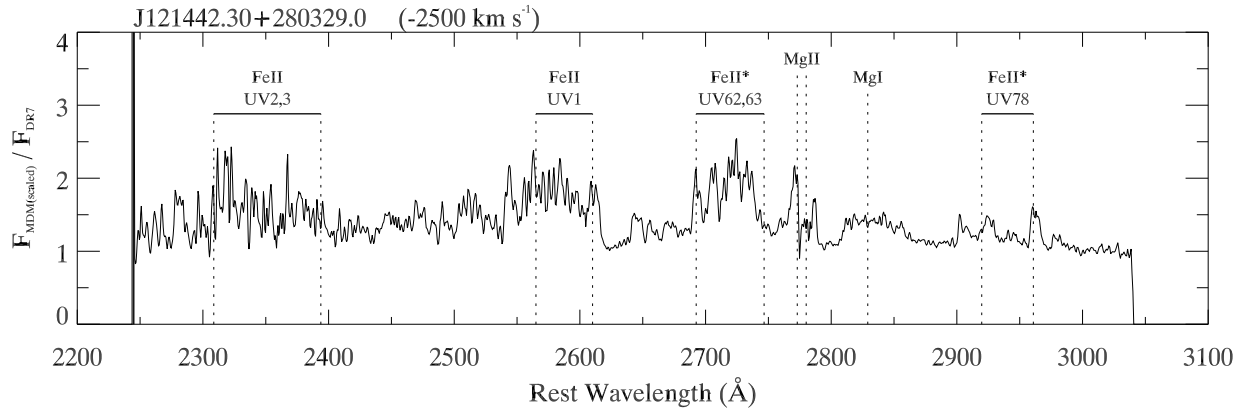}
\caption{Ratio between the scaled 2010 MDM and DR7 spectra for J121442, revealing differing amplitudes of variability from line to line.}
\end{center}
\end{figure*}

In order to conservatively analyze variability due to the Mg\thinspace \textsc{ii} $\lambda\lambda$2796, 2803 BAL, the MDM epoch was rescaled to subtract off the large-scale deviation in the ratio spectrum. The red dashed line in Fig. 3 shows this scale function as it was fit to the ratio spectrum and optimized for wavelengths longward of $\sim$2550 \AA; Fig. 4 shows the resulting rescaled MDM spectrum plotted against the DR7 epoch. Using the rescaled version of the MDM observation, we detect absorption line variability meeting our 8$\sigma$ criterion in three intervals between $\sim$2690--2710 \AA, with the largest variation reaching 9.1$\sigma$. The average width of the variable intervals was $\sim$490 km s$^{-1}$. The Mg\thinspace \textsc{ii} BAL absorbs from approximately 2670 to 2720 \AA \ and likely shortward of 2670 \AA; the lower limit is uncertain due to the onset of possible absorption from Fe\thinspace \textsc{ii}$^*$ (UV62,63).

Although it would be useful to subtract out the large-scale deviations and study the residual variability $\la2550$ \AA, the lack of regions to constrain the fit at the blue end would produce large uncertainties. We conclude that the detected variability of J120337 derives from a unknown change from either the quasar continuum or absorption lines along with weakening of the Mg\thinspace \textsc{ii} $\lambda\lambda$2796, 2803 BAL over a $\sim$3.1 yr time-scale.

\subsection{J121442.30+280329.0}

Observations for J121442 span time-scales from $\sim$15 d to $\sim$7.6 yr between any two spectra. We did not detect significant variability when comparing any two HET epochs from 2000, therefore we created a combined HET spectrum via a weighted average.

The shortest time-scale of absorption line variability meeting the 8$\sigma$ criterion was $\sim$0.6 yr between 2010 and 2011, and the most significant absorption line changes, at $\sim$20$\sigma$ significance, occurred over the $\sim$5.9 yr period between 2000 and 2010. The observed changes can be attributed to the absorption lines because inspection of the ratios shows features that mimic the series of lines in the actual spectrum. These deviations are localized relative to the wavelength coverage and therefore cannot be explained by any intrinsic, continuum variability or systematic issues.

From 2006 to 2010 we see significant decreases in absorption line depths from nearly all of the transitions across the range 2320--2970 \AA \ (see Fig. 1). This pattern met our 8$\sigma$ criterion for transitions involving Fe\thinspace \textsc{ii} (UV1,2,3), Fe\thinspace \textsc{ii}$^*$ (UV62,63,78), Mg\thinspace \textsc{i} $\lambda$2852, and the Mg\thinspace \textsc{ii} $\lambda\lambda$2796, 2803 BAL, although it is uncertain whether the last region from $\sim$2780--2790 \AA \ is due to the Mg\thinspace \textsc{ii} BAL or BEL. We can extract additional information by taking the ratio between the scaled 2010 MDM spectrum and the 2006 DR7 epoch in order to see how the strength of the variability varies from line to line; Fig. 5 shows the resulting ratio. The small regions around 2625 and 3000 \AA \ are approximately unity and thus serve as reference points to where there were no changes between the two epochs. It is interesting that the amplitudes of line ratios within a given Fe\thinspace \textsc{ii} multiplet are not all the same. If we assume that the only contribution to the absorption in these regions is due to the identified Fe\thinspace \textsc{ii} transitions in an optically thin regime then we should expect the line amplitudes within a given multiplet to be the same, independent of the origin of the variability. The fact that they are not implies that there may be extra contributions to the absorption lines due to either higher excitation state transitions in Fe\thinspace \textsc{ii} or other ionic species, or that some Fe\thinspace \textsc{ii} transitions are becoming saturated over our measured time-scales.

Over the next $\sim$0.6 yr from 2010 to 2011 there are intervals of variability ranging from $\sim$2610--2960 \AA, which include Fe\thinspace \textsc{ii} (UV1), Fe\thinspace \textsc{ii}$^*$ (UV62,63,78), the Mg\thinspace \textsc{ii} $\lambda\lambda$2796, 2803 BAL--BEL, and Mg\thinspace \textsc{i} $\lambda$2852. The absorption line depth is increasing for these intervals, opposite to what happened in the 2.3 yr before. There is no statistically significant variability over the $\sim$0.5 yr from 2011 to 2012 or the $\sim$0.6 yr from 2012 to 2013, although there is evidence of the continuing trend of increasing absorption line depth over these time-scales\footnote{This trend is shown visually in Fig. A (see additional online material).}
. When comparing the 2010 and 2013 epochs, we detect absorption line variability with the 8$\sigma$ criterion   in wavelength intervals ranging from $\sim$2390--2970 \AA.

We conclude that across the $\sim$7.6 yr baseline for J121442 the absorption line variability is complex. The absorption lines associated with Fe\thinspace \textsc{ii} and Mg\thinspace \textsc{i} $\lambda$2852 exhibit large, significant decreases in strength over 2.3 yr (see Fig. 1), followed by increases in depth over the next 1.8 yr (see Fig. A).

\subsection{J123103.71+392903.7}

We have a time baseline spanning $\sim$0.5--3.9 yr for J123103 with four observations for comparisons. Fig. B (see additional online material) shows comparisons between the DR7 epoch and the other three available spectra. No absorption line variability was detected with 8$\sigma$ confidence between any two spectra from 2011--2013. The comparison between the DR7 and DR12 epochs in Fig. 1 show that variations occur from $\sim$1900--3200 \AA, but do not exist $\ga$ 3200 \AA. The coincidence between the identified ionic species with the variability in Fig. 1 is evidence that the observed changes are due to the absorption lines and not intrinsic in nature or due to systematic errors. We note that, in addition to the lines in the template, other multiplets of Fe\thinspace \textsc{ii} and other ions are likely present to contribute to the absorption from $\sim$1900--3200 \AA.

Over the $\sim$2.9 yr between the DR7 and DR12 epochs, we see decreases in absorption line strength using our 8$\sigma$ criterion between $\sim$2130 and 2690 \AA, a region where the Fe\thinspace \textsc{ii} (UV1,2,3), Fe\thinspace \textsc{ii}$^*$ (UV62,63), and Ni\thinspace \textsc{ii}$^*$ (UV12,13) multiplets occur (see Fig. 1). This source varied with a maximum significance level of 11$\sigma$. Similar decreases in absorption line strength are observed near Fe\thinspace \textsc{iii}$^*$ (UV48) and Fe\thinspace \textsc{ii}$^*$ (UV78,Opt6), although these do not meet our 8$\sigma$ threshold. The BAL extending from $\sim$2650 to 2775 \AA \ could consist of absorption from both Mg\thinspace \textsc{ii} and Fe\thinspace \textsc{ii}$^*$ (UV62,63), and we see variability only in the high-velocity portion of the BAL.  

We conclude that there is no significant absorption line variability from 2011 to 2013, a time-scale of $\sim$1.0 yr, however J123103 does exhibit absorption line weakening from 2005 to 2011 at the 8$\sigma$ level. 

\subsection{Tentative sources}

Sources that meet the $6\sigma$ threshold for tentative variability over wavelength intervals associated with absorption (J104459, J112220, J112526) may represent real changes that do not meet our $8\sigma$ threshold. In order to further assess the variability in these objects, we conducted tests using larger bin widths than the 280 km s$^{-1}$ width used in the general analysis and inspected comparisons shown in Fig. 1; the larger bin widths probe variations over entire BAL troughs, providing additional information beyond the general analysis which is optimized to study variability over small portions of BALs.

J104459 shows increases in absorption line amplitude centered at $\sim$2650 \AA \ and 2700 \AA, although the former wavelength is the only interval that reaches $6\sigma$ (see Fig. 1). Tests involving larger bin widths yield a maximum significance level of $\sim$19$\sigma$ using a bin width of 10500 km s$^{-1}$, corresponding to $\sim$70 per cent of the width of the absorption feature at $\sim$2675 \AA. Because the variations are localized relative to the wavelength coverage and no other intervals show variability, the tentative variations in this source are likely real.  The absorption may correspond to a high velocity component of Mg\thinspace \textsc{ii}.

The degree of variability in J112220 is significant up to $\sim$28$\sigma$, using a bin width of $\sim$10850 km s$^{-1}$ centered on 2590 \AA \ (see Fig. 1); J112526 exhibits 8$\sigma$ variations that may be due to the Mg\thinspace \textsc{ii} BEL and we therefore could not obtain a reliable estimate of the tentative absorption variability. Both J112220 and J112526 do not exhibit localized variability but show changes scattered across a large portion of the wavelength coverage. These variations may be due to  Fe\thinspace \textsc{ii} BELs instead of absorption lines and the reality of the absorption variability in these two sources is therefore less likely.

J142703 exhibits 8$\sigma$ variations from $\sim$2480--2510 \AA, a localized region that is not associated with any transition from the line template (see Fig. 1). The variability is unlikely a flux calibration issue since J121442 was observed on the same night as J142703 and does not exhibit similar behavior. Variability due to Fe \thinspace \textsc{II} BELs is also unlikely since we would expect changes over larger wavelength intervals. The observed variation is therefore potentially due to absorption and might be associated with highly blue-shifted Mg\thinspace \textsc{ii} $\lambda\lambda$2796, 2803. J084044 is not included in this analysis due to the ambiguity in associating variations with absorption features.

We conclude that, of the sources categorized as tentative, J104459 and J142703 exhibit absorption line variability that is likely real and associated with blue-shifted Mg\thinspace \textsc{ii} $\lambda\lambda$2796,2803.

\subsection{Associated absorption lines}

Fig. 2 shows spectra comparisons of the Mg\thinspace \textsc{ii} $\lambda\lambda$2796, 2803 and Mg\thinspace \textsc{i} $\lambda$2852 AALs for J030000 and J123103, the only two sources in our sample that have AALs. J030000 did not show variability under our 6$\sigma$ criterion between any of the three DR7 observations; we therefore used the weighted average epoch to compare with the MDM observation, a time-scale of $\sim$6.0 yr. 

We detect tentative variability in J030000 meeting the 4$\sigma$ criterion in the Mg\thinspace \textsc{ii} $\lambda$2803 AAL and also at $\sim$2815 \AA, which may be associated with red-shifted Mg\thinspace \textsc{ii} or some other unidentified ion (see Fig. 2). Both of the Mg\thinspace \textsc{ii} and Mg\thinspace \textsc{i} AALs increase in amplitude upon inspection by eye, however the variations occur across $\sim$2 pixels, making it hard to consider the changes as real. It is interesting to note, however, that the Mg\thinspace \textsc{ii} AALs are at equal depths but do not reach zero flux, an indication that Mg\thinspace \textsc{ii} is saturated and partially covers the continuum source. Using our 6$\sigma$ threshold, we conclude that there is no significant AAL variability in J030000.

We also did not detect AAL variability at 6$\sigma$ between the four observations of J123103, with time-scales up to $\sim$3.9 yr. We note, however, that the comparison between 2011 and 2013 (see Fig. 2) shows tentative changes at 4$\sigma$ in the Mg\thinspace \textsc{i} $\lambda$2852 line. The amplitude of the Mg\thinspace \textsc{i} AAL appears to increase over the $\sim$1.0 yr time-scale and the variation persists over multiple pixels, however we conclude that the changes are only marginal.

The lines we detect in J030000 and J123103 are likely composed of narrower, unresolved features; average measured FWHM for the Mg\thinspace \textsc{ii} and Mg\thinspace \textsc{i} AALs are $\sim$150 and 135 km s$^{-1}$ for J030000, and $\sim$270 and 210 km s$^{-1}$ for J123103, respectively. We note that a portion of the AAL profiles in J123103 is associated with blue-shifted absorption, while the rest appears to be red-shifted. This behavior, combined with the potential red-shifted Mg\thinspace \textsc{ii} AAL system in J030000, indicates that the AAL gas is complex with possible outflow and inflow components. We conclude that there is no AAL variability meeting our 6$\sigma$ criterion in either of the two sources analyzed, however future observations may provide insight into the nature of the possible variations.

\section{Discussion}

\subsection{Interpreting the variability}

Placing constraints on outflows requires an understanding of the physical origin of the absorption line changes. Two scenarios that can potentially explain the variability are that either the absorbing gas undergoes an ionization change or a change in covering fraction occurs along our LOS due to transverse motions of gas across the emission region of the AGN\footnote{A variant of the second scenario is that a portion of the continuum source outside the absorber sightline undergoes a local change in emissivity, leading to fluctuations in an unabsorbed continuum component.}. Below we discuss the origin of the variations on the basis of photometric and optical depth arguments, and provide evidence that broadly supports change in covering fraction as the preferred scenario to explain absorption line variability in our sample.
 
 \subsubsection{Photometric arguments}
 
The photometric behavior of our sources may provide information regarding the origin of the absorption line variations. We utilize synthesized CSS $V$-band photometry to look for significant flux variations over our measured time-scales. Fig. 6 shows photometry of the three Y sources along with spectral observation dates from our sample (red dotted lines)\footnote{Photometric changes in the other sources in our sample were found to be $\la$ 0.1 mag when comparing CSS measurements at the epochs for our spectroscopic observations.}. There is a small decrease ($\sim$0.16) in $V$-band magnitude between the two observations in J120337; a decrease of $\sim$0.18 mag is observed between the first two epochs in J121442, which is followed by a smaller increase from $\sim$2010--2012. Photometry provided for J123103 is less reliable due to large variations that likely originate from systematic errors (the source is close to a bright star). 

Comparisons between our spectroscopic findings and measurements in Fig. 6 reveal a noteworthy result in that absorption line strength correlates inversely with estimated $V$-band flux. This behavior is seen for each of the epochs in J120337 and J121442 and possibly in J123103. A qualitative pattern of this type is expected in scenarios involving gas absorbing in the $V$-band and crossing the LOS while the background light source remains constant. Alternatively, the $V$-band flux primarily traces changes in the source continuum luminosity, with the absorber ionization state and ion column density responding to the incident radiation field to produce the observed trend. Since we do not have absolute photometry for most spectra, it is ambiguous whether the background light source changed in our objects. However, a calculation of the fractional change in synthesized $V$-band flux for our scaled spectra can be compared with the CSS measurements to constrain the intrinsic continuum variability.

The spectra shown in Fig. 1 would thus predict for J120337 a $\sim$40 per cent increase in synthesized $V$-band flux between 2005 and 2012; the CSS measurements show a much smaller increase of $\sim$13 per cent. The difference can be understood if the weakening of absorption is accompanied by a drop in continuum luminosity, however inspection of Fig. 1 suggests an alternative explanation of a change in continuum shape or possibly systematic errors. For J121442, our wavelength coverage does not allow us to synthesize a $V$-band flux, however using spectra from 2006--2010 we compute a $\sim$23 per cent increase in flux over 3800--5150 \AA; this is slightly larger than the $\sim$15 per cent change derived using CSS fluxes near those dates. If the modulations are only due to absorption and variable coverage, we would expect higher amplitudes at shorter wavelengths in this source due to the larger number of transitions, consistent with this comparison. Our third source, J123103, yields a $\sim$20 per cent increase in $V$-band flux between 2005 and 2011, calculated using spectra from our sample. CSS photometry for J123103 has systematic uncertainties and therefore a comparison could not be tested quantitatively, however there is also evidence of a flux increase from 2005--2011. 

In summary, with the current data we cannot distinguish on the basis of photometric arguments whether variability is more likely due to a change in coverage or an ionization change.

   \begin{figure}
\begin{center}
\includegraphics[width=1.0\columnwidth,angle=0]{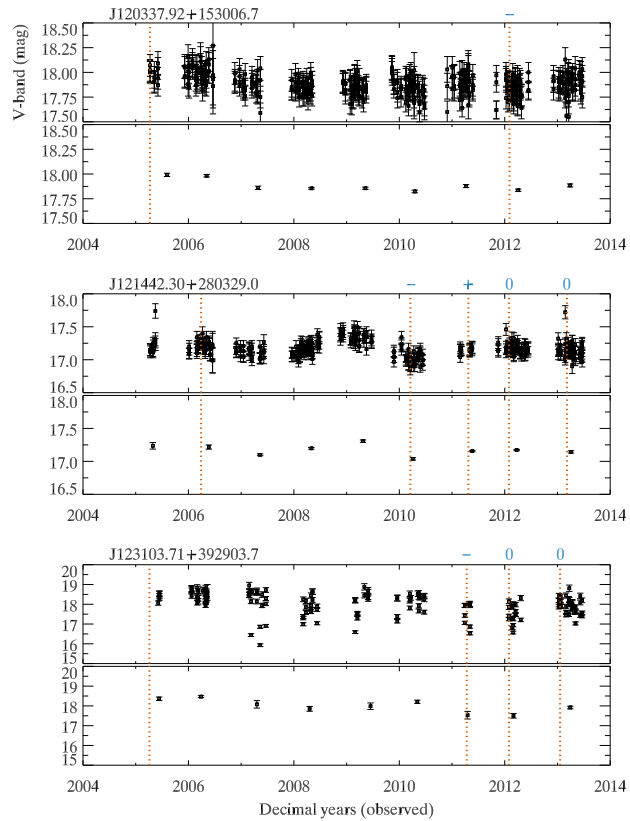}
\caption{Estimated CSS $V$-band photometry for the Y sources, depicting variations between spectral observations (shown as red dotted lines). Each upper panel shows all photometric measurements with their errors, and each lower panel shows average measurements over one year increments. Blue symbols above the red dotted lines indicate the directional change in absorption line amplitude relative to the previous spectral observation; absorption lines either decreased (--), increased (+), or did not meet our 8$\sigma$ criterion (0).}
\end{center}
\end{figure}
 
 \subsubsection{Optical depth arguments}
 
Variability in absorption lines that are known to be optically thick, if observed, provides strong evidence for changes in coverage (e.g. Capellupo et al. 2014a). The Mg\thinspace \textsc{ii} feature can potentially be used as such a diagnostic, and information on its optical depth can be derived from considering the combined behavior of Mg\thinspace \textsc{ii} and Fe\thinspace \textsc{ii} absorption.
 
For that purpose, we calculated a different optical depth $\tau_{\rmn{Fe}}$ for each Fe\thinspace \textsc{ii} multiplet present in our sample (UV1,2,3,62,63,78,Opt6) and use those values with the optical depth $\tau_{\rmn{Mg}}$, calculated using the Mg\thinspace \textsc{ii} $\lambda\lambda2796, 2803$ doublet, to determine a set of ratios  $\tau_{\rmn{Mg}}/\tau_{\rmn{Fe}}$ that depend on the respective ionic column densities, transition oscillator strengths, and transition wavelengths. For each ratio we adopted average transition wavelengths and oscillator strengths from \citet{ver96} and \citet{fuh06}. 
 
 For our calculations, we assumed the Mg\thinspace \textsc{ii} to Fe\thinspace \textsc{ii} column density ratio to be equal to the solar elemental abundance ratio (taken from \citealt{asp09}). We did not factor in the unknown level populations of Fe\thinspace \textsc{ii} and Mg\thinspace \textsc{ii} or the distribution of iron and magnesium over their ionization stages, but adopted the solar abundance ratio throughout. Incorporating details about the physical conditions of the gas, however, only favors larger  $\tau_{\rmn{Mg}}/\tau_{\rmn{Fe}}$ than the ones calculated here due to the large number of low-lying Fe\thinspace \textsc{ii} energy levels and closely spaced ionization potentials in iron relative to magnesium. Since we are only concerned with establishing very crude lower limits on  $\tau_{\rmn{Mg}}/\tau_{\rmn{Fe}}$ for the purpose of understanding the variability, the above assumptions do not affect our interpretations. 
 
We calculate lower limits on  $\tau_{\rmn{Mg}}/\tau_{\rmn{Fe}}$ to be, on average, $\sim$6.3 using Fe\thinspace \textsc{ii} (UV1,2,3) multiplets, $\sim$4.1 using Fe\thinspace \textsc{ii}$^*$ (UV62,63), $\sim$15 using Fe\thinspace \textsc{ii}$^*$ (UV78), and $\sim$50 for Fe\thinspace \textsc{ii}$^*$ (Opt6). Observing Fe\thinspace \textsc{ii} in absorption at a given optical depth therefore implies that the Mg\thinspace \textsc{ii} optical depths can range from being at least $\sim$4.1--50 times larger than Fe\thinspace \textsc{ii}, depending on which Fe\thinspace \textsc{ii} multiplets exhibit absorption. We use this information below to assess whether a change in covering fraction due to transverse motions of gas is the likely scenario occurring in the three Y sources.
 
J120337 appears to exhibit absorption from Fe\thinspace \textsc{ii}$^*$ (UV62,63) at low optical depths but no detectable absorption from Fe\thinspace \textsc{ii}$^*$ (UV78,Opt6), and Mg\thinspace \textsc{ii} is therefore less likely to be saturated (see Fig. 1). We concluded in \S 4.1 that the observed variability in this source is due to both an unknown change across the entire wavelength coverage and a weakening in the Mg\thinspace \textsc{ii} $\lambda\lambda$2796,2803 BAL. Localized changes in the ratio spectrum are detectable near transitions from Fe\thinspace \textsc{ii} (UV1) and Fe\thinspace \textsc{iii}$^*$ (UV48), implying that these lines may have varied in the same manner as Mg\thinspace \textsc{ii} (see Fig. 3). However, the poor signal-to-noise ratio towards the blue end and the fact that we cannot fit these regions with any degree of certainty makes it inconclusive if this behavior is occurring in a significant way. Ions of different ionization state varying in concert supports the scenario of a change in covering fraction due to transverse motions of gas, however the fact that Fe\thinspace \textsc{iii} has a ground state ionization potential not too dissimilar from Fe\thinspace \textsc{ii} ($\sim$31 and 16 eV respectively, \citealt{dra11}) makes it difficult to rule out an ionization change in this source.
 
J121442 shows detectable absorption coming from Fe\thinspace \textsc{ii} (UV1,2,3), Fe\thinspace \textsc{ii}$^*$ (UV62,63,78), and Fe\thinspace \textsc{ii}$^*$ (Opt6); this implies that Mg\thinspace \textsc{ii} is very likely saturated in this source (see Fig. 1). Significant variability meeting the 8$\sigma$ criterion occurred in regions associated with Fe\thinspace \textsc{ii} (UV1), Fe\thinspace \textsc{ii}$^*$ (UV62,63,78), and the Mg\thinspace \textsc{ii} $\lambda\lambda$2796, 2803 BAL (see Fig. 1), although the variable interval near Mg\thinspace \textsc{ii} may be due to either the BAL or BEL. The lack of variability in the deep portion of the Mg\thinspace \textsc{ii} BAL might be due to differing covering fractions between the Mg\thinspace \textsc{ii} and Fe\thinspace \textsc{ii} gas in an inhomogeneous absorber; large $\tau_{\rmn{Mg}}/\tau_{\rmn{Fe}}$ ratios along with closely spaced ionization potentials in iron relative to magnesium support a higher Mg\thinspace \textsc{ii} covering fraction than Fe\thinspace \textsc{ii} (see \citealt{luc14}). Futhermore, \citet{dek02a} used a template-fitting approach to model the spectrum of J121442 and constrained the Fe\thinspace \textsc{ii} column density to be $\sim$10$^{16.95\pm0.10}$ cm$^{-2}$. They concluded that the high column densities imply that Fe\thinspace \textsc{ii} is saturated and thus partially covers the continuum source; Fig. 1 shows evidence of this as the regions where the Fe\thinspace \textsc{ii} (UV1,2,3) multiplets exist are approximately flat and do not reach zero flux. The likely interpretation to explain the variability in J121442 is therefore a change in covering fraction from transverse motions of the outflowing gas. 

Mg\thinspace \textsc{ii} is also likely saturated in J123103, as this source exhibits absorption coming from Fe\thinspace \textsc{ii} (UV1,2,3) and Fe\thinspace \textsc{ii}$^*$ (UV62,63,78), with the possible presence of Fe\thinspace \textsc{ii}$^*$ (Opt6) at very small optical depths. Significant decreases in absorption line strength meeting the 8$\sigma$ criterion are observed in the regions with transitions from the Fe\thinspace \textsc{ii} (UV1,2,3) and Fe\thinspace \textsc{ii}$^*$ (UV62,63) multiplets; no variability is detected in the lower velocity portion of the Fe\thinspace \textsc{ii}$^*$ (UV62,63)--Mg\thinspace \textsc{ii} $\lambda\lambda$2796, 2803 BAL trough (see Fig. 1). The same changes are observed near Fe\thinspace \textsc{iii}$^*$ (UV48) and Fe\thinspace \textsc{ii}$^*$ (UV78,Opt6), although these regions did not meet our 8$\sigma$ criterion.

 An ionization change within the gas that ionizes Fe\thinspace \textsc{ii} to Fe\thinspace \textsc{iii} would predict weakening in Fe\thinspace \textsc{ii} absorption lines and strengthening in the Fe\thinspace \textsc{iii}, which is not observed. It is possible that both ions are decreasing in favor of higher ionization stages. It is noteworthy, however, that only a portion of the Fe\thinspace \textsc{iii} trough weakens. Detecting variability in portions of BALs can be explained by a change of covering fractions due to the observed ionic species moving out of the line of sight only at the higher velocities; evidence of saturation in the Fe\thinspace \textsc{ii} (UV1,2,3) and Ni\thinspace \textsc{ii} (UV12,13) lines also lends support to transverse motions of gas in J123103, as the wavelength intervals across these multiplets are approximately flat and do not reach zero flux (see Fig. 1). If Mg\thinspace \textsc{ii} is saturated, however, the lack of variability in the lower velocity portion of the BAL around 2750 \AA \ is also consistent with an ionization change scenario. It is possible that this source is undergoing both gas motion across the LOS and ionization changes; Fig. 1 also shows weakening in the Fe\thinspace \textsc{ii}$^*$ (UV78) line across its entire absorbing range. 
 
 In light of the above evidence for each of our Y sources, the observed changes are readily interpreted as a change in covering fraction, and this explanation is preferred in particular for J121442 and J123103. We therefore consider an outflow crossing the LOS in order to place physical constraints on the outflows in our sample. 

 \subsection{Constraining FeLoBAL outflows}
 
 Evidence favoring transverse motions of gas across the LOS to explain the observed variability allows us to constrain the distance of the absorbing ions from the SMBH using the measured variability time-scale $\Delta t$, an estimate of the size of the continuum region at a particular wavelength, $D_{\lambda_{\rmn{peak}}}$, and the approximate fraction $\Delta A$ of the continuum region that the outflow traversed over the measured time-scale. Following \citet{cap11}, we quantify $\Delta A$ by measuring the average flux of a variable region for two epochs of interest ($f_1$ and $f_2$) along with the average continuum flux $f_{\rmn{c}}$ across the same interval; we then calculate the ratio $(f_{1}-f_{2})/f_{\rmn{c}}$ to serve as the estimate of $\Delta A$. Due to the many blended BALs present across our sample, determining where to place the continuum was difficult and highly uncertain. We therefore conducted multiple tests, placing the continuum at extreme locations that we believe should encompass the range of possible continuum values, and quantified average estimates on $\Delta A$ and its associated error for each source. Our estimates for $\Delta A$ were $0.21\pm0.10$, $0.19\pm0.11$, and $0.34\pm0.18$ for J120337, J121442, and J123103, respectively. 
 
 For an estimate of $D_{\lambda_{\rmn{peak}}}$, we utilize the result from Chen et al. (in preparation) who assume a standard thin accretion disk that emits as a blackbody and use Wien's displacement law to translate the temperature at a given radius to the wavelength corresponding to maximum blackbody emission to estimate the radius of the disk at that wavelength. Their result adopts a modified constant in the radius function derived using the above assumptions; the new factor is meant to compromise between this standard analysis, \citet{bla11} who used micro-lensed quasars to estimate accretion disk sizes, and \citet{gas07} who argued that the temperature profiles of real AGN disks are flatter than the standard predictions. 
 
 The calculations of $D_{\lambda_{\rmn{peak}}}$ require estimates of the Eddington ratio $L_{\rmn{bol}}/L_{\rmn{edd}}$, SMBH mass $M_{\bullet}$, radiative efficiency $\eta$, and maximum blackbody emission wavelength $\lambda_{\rmn{peak}}$. We took values for $L_{\rmn{bol}}/L_{\rmn{edd}}$ and $M_{\bullet}$ from \citet{she11} to be $\sim$0.06 and $7 \times 10^{9} M_{\sun}$ for J120337, $\sim$0.07 and $3 \times 10^{9} M_{\sun}$ for J121442, and $\sim$0.03 and $4 \times 10^{9} M_{\sun}$ for J123103\footnote{Single epoch estimates of $M_{\bullet}$ for our Y sources were based on either the Mg\thinspace \textsc{ii} (J120337, J123103) or H $\beta$ (J121442) BEL.}. We did not use their quoted statistical errors on $L_{\rmn{bol}}$ and $M_{\bullet}$ for our calculations; we instead adopt uncertainties based on estimates by \citet{she13} to be $\sim$0.3 and 0.5 dex for $L_{\rmn{bol}}$ and $M_{\bullet}$, respectively, reflecting systematic errors associated with single epoch measurements (see \citealt{she13} for a review of quasar black hole mass measurement methods and discussion of these uncertainties)\footnote{Note in particular that FeLoBAL spectra may typically be more reddened than the rest of the quasar population (e.g. \citealt{spr92}); therefore bolometric corrections derived by \citet{she11} may not be representative of these sources and can lead to systematic errors in $L_{\rmn{bol}}$ and $M_{\bullet}$.}. We adopt 0.10 as our estimate of $\eta$, and calculate an average value of $\lambda_{\rmn{peak}}$ to be $\sim$2700 \AA \, based upon the location of the variable regions in the three Y sources.
 
 Using the estimates for the respective continuum source sizes, we consider the extreme case of a circular outflow of infinite radius crossing a square continuum (the so-called `knife-edge' model taken from \citealt{cap13}, see their fig. 14b). We follow the calculation of \citet{cap13} with this model to determine the crossing speed to be $v_{\rmn{cross}}=\Delta A\times D_{\lambda_{\rmn{peak}}}/ \Delta t$. We can then tie the crossing speed to a distance assuming that the outflow is in keplerian rotation around the SMBH (i.e. $r=GM_{\bullet}/v_{\rmn{cross}}^{2}$).  The knife-edge model presented by \citet{cap13} is meant to be a limiting case of all geometrical situations involving an outflow traversing our LOS to the continuum-emitting accretion disk, thereby establishing a lower limit on $v_{\rmn{cross}}$ and upper limit on the distance $r$ from the SMBH. 
 
Using the above parameters and their errors, we estimate diameters of the continuum emitting regions for J120337, J121442, and J123103 to be $\sim$0.031$^{+0.016}_{-0.012}$, $0.018^{+0.010}_{-0.007}$, and $0.016^{+0.009}_{-0.006}$ pc, respectively; the errors were propagated through using the estimated uncertainties on $L_{\rmn{bol}}$ and $M_{\bullet}$ from \citet{she13}. Using our respective estimates on $\Delta A$ and their errors, we constrain lower limits on the crossing speeds to be $v_{\rmn{cross}}\ga550$, $990$, and $540$ km s$^{-1}$ for J120337, J121442, and J123103, respectively. These yielded upper limits on the outflow distance from the SMBH to be $r\la26$, $2.5$, and $23$ pc for our three Y sources, respectively.

The upper limits on $r$ do not account for errors on $\eta$ and $\lambda_{\rmn{peak}}$. While $\lambda_{\rmn{peak}}$ can only fluctuate on the order of tens of angstroms over our variable regions, $\eta$ can reach up to 0.42 (for a maximally spinning Kerr black hole in co-rotation with the accretion disk). Our adopted $\eta$ $\sim$ 0.1 estimate is potentially uncertain, given that recent work has estimated high spins and radiative efficiencies on massive black holes in quasars from $z$ $\sim$ 1.5--3.5 \citep{tra14,cap15}. Since we do not have information about the spins of the SMBHs in our sample, we consider the extreme case by adopting $\eta$ to be 0.42 and $\lambda_{\rmn{peak}}$ to be the shortest wavelength in each variable region used; these extreme parameters yield the maximum possible upper limit on $r$ for each of our variable sources. When also accounting for errors from the other parameters, we thus derive conservative upper limits to be $r\la 69$, $7$, and $60$ pc for J120337, J121442, and J123103, respectively.

\subsection{Comparisons to previous work}

Our measured variability time-scales of $\sim$0.6--7.0 yr and our limits on outflow distance for the Y sources are broadly consistent with previous variability studies of FeLoBAL quasars. \citet{viv12} saw variable time-scales down to $\sim$1.2 yr in one out of the five FeLoBAL quasars in their sample, and attributed the changes to Fe\thinspace \textsc{iii} and Al\thinspace \textsc{iii} fine-structure lines. \citet{hal11} detected absorption line variability over 0.6--5 yr in the FeLoBAL quasar J1408+3054, observing dramatic changes in Mg\thinspace \textsc{ii} and disappearance of Fe\thinspace \textsc{ii} lines at the same velocity; using a different outflow model than the one adopted in our analysis, \citet{hal11} constrained the absorber distance to be between 1.7 and 14 pc from the SMBH, in approximate accord with our results. 

The general consensus from previous work on HiBAL quasars is that BAL variability is common over multiple year time-scales (see \citealt{fil13}, \citealt{cap13}, and refs. therein). This is consistent with our results. Previous studies on C\thinspace \textsc{iv} and Si\thinspace \textsc{iv} BAL variability have also concluded that variable regions tend to occur in portions of BAL troughs (\citealt{gib08,cap12,cap13,fil13}). \citet{fil13} calculated the mean velocity width of variable regions across their sample of C\thinspace \textsc{iv} and Si\thinspace \textsc{iv} BALs to be 713.6 km s$^{-1}$ and 592.8 km s$^{-1}$, respectively. These are on the same order of magnitude as our Mg\thinspace \textsc{ii} BAL estimate of $\sim$520 km s$^{-1}$ for our three Y sources. \citet{cap11,cap12,cap13} constrain their sample of variable C\thinspace \textsc{iv} and Si\thinspace \textsc{iv} BAL outflows to distances $\la 10$ pc.

Previous work that has constrained quasar outflow distances using photoionization modeling yields a range of results. \citet{dun10} studied J031856, categorized as a `bright, reddened' FeLoBAL quasar by \citet{hal02b},  and divided the BALs into approximately eleven separate kinematic components. \citet{dun10} also determined the electron density of the gas using metastable excited levels from Si\thinspace \textsc{ii}, Fe\thinspace \textsc{ii}, and Ni\thinspace \textsc{ii} and, combined with an estimate of $L_{\rmn{bol}}$, constrained the distance of the outflowing gas to likely be $\sim$6 kpc from the SMBH. Our lack of variability in J031856 with our available data is consistent with the distance estimate of \citet{dun10}; photoionized gas at larger distances from the SMBH implies lower densities (for a given ionization parameter) which result in longer recombination time-scales for an ionization change to occur, and transverse motions across the line of sight are less likely to produce detectable changes from absorbing ions farther away from the SMBH. 
     
\citet{ara13} used density-sensitive, unblended absorption lines from O\thinspace \textsc{iv} in the far UV spectrum of the non-BAL quasar HE0238-1904 to estimate an outflow distance of 3 kpc from the SMBH. \citet{dek01} analyzed the FeLoBAL quasar J104459 by assuming a single-phase outflow model and determined the distance of the Fe\thinspace \textsc{ii} and Mg\thinspace \textsc{i} absorbing region from the SMBH to be $\sim$700 pc. \citet{eve02} conducted a follow-up analysis of this source, instead assuming a shielded, multi-phase outflow that consisted of high-density clouds embedded in a low-density, continuous magnetohydrodynamic wind. This reinterpretation to explain the observed nature of the absorption lines in J104459 yielded a new distance estimate between the absorbing gas and SMBH to be $\sim$4 pc. These two estimates are model dependent and differ by approximately two orders of magnitude. Over our available time-scale of $\sim$4.0 yr we detect tentative absorption line variability with our 6$\sigma$ criterion for J104459. It is interesting that the tentative variations are likely from a localized region that could be highly blue-shifted Mg\thinspace \textsc{ii}, while the other absorption line systems do not vary. Our results are therefore broadly consistent with both \citet{dek01} and Everett et al. (2002) in the sense that this source may have outflowing gas at two distinct radii.

Order of magnitude differences in the distances of outflows from the SMBH have direct implications for theoretical models attempting to explain the outflow origin and underlying acceleration mechanisms. Quasar outflows have been modeled as winds arising in the same region as the BELs on sub-parsec scales (see e.g., \citealt{mur95}, \citealt{pro00}, \citealt{pro04}, \citealt{fuk10}), while Faucher-Gigu\`ere et al. (2012) explained FeLoBAL formation in particular to be from the quasar wind propagating out to kiloparsec scales and interacting with ISM gas to form clumps that produce the observed absorption (see also \citealt{fau12a}). 

\citet{luc14} discussed the selection bias associated with photoionization models that require unblended lines from both ground and excited states of a given ion to constrain the density of the outflowing gas; they argue that density-sensitive lines observed in BAL quasar spectra are only useful in lower density gas and outflow distance constraints are therefore larger. Looking for objects with these characteristics may inherently bias the result and subsequent conclusions may not extend to the entire population of outflows. \citet{luc14} estimated the outflow in the FeLoBAL quasar FBQS J1151+3822 to be between 7.2 and 127 pc from the SMBH using photoionization modeling and blended, excited state lines in Fe\thinspace \textsc{ii} and He\thinspace \textsc{i}, a range that is similar to results from variability studies. 

Our upper limit of $\sim$7 pc for J121442 is consistent with the results of \citet{dek02a}, who studied a Keck HIRES spectrum of this source by fitting a model to the observed data in order to extract column densities, excitation temperature, and additional information on the continuum and covering factor of the outflow. In conjunction with photoionization models, \citet{dek02a} were able to constrain the distance of the outflow in J121442 to be between 1 and 30 pc from the SMBH. These distance constraints should also broadly apply to other quasars that have similar spectra as J121442 (i.e., large Fe\thinspace \textsc{ii} column densities and the presence of Fe\thinspace \textsc{ii}$^*$ absorption lines); sources in our sample that have similar luminosities and absorption spectra, and therefore are approximately consistent with the $1-30$ pc distance constraint of \citet{dek02a}, include J030000, J084044, J112220, J112526, J115436, and J123103 (see Fig. 1).  

Other studies have found a range of outflow distances to exist in the same object. \citet{dek02b} analyzed a Keck HIRES spectrum of J084044, a `radio--moderate' BAL quasar \citep{bec97}, concluding that there are two distinct kinematic outflow components with different densities; the distances of the high and low-density absorbing regions from the SMBH were found to be $\sim$1 and 230 pc, respectively, based on column density measurements of Fe\thinspace \textsc{iii} and Al\thinspace \textsc{iii} along with level populations observed in low-lying states in Ni\thinspace \textsc{ii}, Si\thinspace \textsc{ii}, and Fe\thinspace \textsc{ii}. \citet{moe09} analyzed the FeLoBAL quasar J0838+2955, constraining the lower velocity BAL system to be at 3.3 kpc from the SMBH using Si\thinspace \textsc{ii}$^*$ and Fe\thinspace \textsc{ii}$^*$ lines with photoionization modeling but also detected variability in the higher velocity C\thinspace \textsc{iv} BAL system and constrained it to be $\la0.1$ pc from the SMBH. \citet{fil13} detected both variable and stable C\thinspace \textsc{iv} troughs over their measured time-scales in two sources (J141513.99+365412.2 and J225608.48+010557.7, see their \S 5.1). In light of the above discussion, it seems plausible that FeLoBAL outflows can exist at a wide range of distances from the SMBH.

\section{Summary}

We have detected absorption line variability using our $8\sigma$ criterion in 3 out of the 12 FeLoBAL quasars in our sample over rest frame time-scales of $\sim$0.6--7.6 yr (J120337, J121442, J123103).  We have also further quantified and assessed the significance of variability in our tentative (T) sources, concluding that two objects likely exhibit real variations in the absorbing region (J104459, J142703).  We have also analyzed Mg\thinspace \textsc{ii} and Mg\thinspace \textsc{i} AAL variability in two of our sources (J030000, J123103), finding only marginal variability using a tentative $4\sigma$ criterion over measured time-scales up to $\sim$3.9 yr.

Wavelength intervals of variability that met our 8$\sigma$ criterion are associated with transitions coming from ground and excited states in Fe\thinspace \textsc{ii} as well as Mg\thinspace \textsc{ii} $\lambda\lambda$2796, 2803, Mg\thinspace \textsc{i} $\lambda$2852, and Ni\thinspace \textsc{ii}$^*$ (UV12,13).  Due to our limited wavelength coverage and signal-to-noise ratio, we were unable to probe variability of Fe\thinspace \textsc{iii} and higher ionization species such as C\thinspace \textsc{iv} and Si\thinspace \textsc{iv}.  Studying a larger range of ionization states simultaneously would be useful in the future for interpreting the origin of the observed variability.

Sources meeting our 8$\sigma$ threshold (J120337, J121442, J123103) exhibit absorption line changes readily interpreted as transverse motions of gas across the LOS; this scenario is preferred in particular for J121442 and J123103 due to evidence of saturation in the Fe\thinspace \textsc{ii}, Ni\thinspace \textsc{ii}, and Mg\thinspace \textsc{ii} absorption lines. Other scenarios cannot be entirely ruled out.  In the crossing absorber scenario, we constrain the outflow distances from the SMBH to be $\la$ 69, 7, and 60 pc for J120337, J121442, and J123103, respectively, with conservative assumptions.

Our results differ by a few orders of magnitude with some studies that utilize photoionization models and constrain outflows to be at kiloparsec distances from the SMBH.  Two possibilities are that either two outflow phases of different origin are being probed, or we are tracing the same outflowing gas in different stages in its evolution.  Determining the incidence of absorbing outflows as a function of distance from the SMBH using large sample studies will be important for understanding this possible connection. 

\section*{Acknowledgements}

FWH acknowledges support from the U.S. National Science Foundation grant AST-1009628. SCG acknowledges support from the Natural Science and Engineering Research Council of Canada. WNB acknowledges support from the U.S. National Science Foundation grant AST-1108604.

This work is based on observations obtained at the MDM Observatory, operated by Dartmouth College, Columbia University, Ohio State University, Ohio University, and the University of Michigan.

Funding for SDSS-III has been provided by the Alfred P. Sloan Foundation, the Participating Institutions, the National Science Foundation, and the U.S. Department of Energy Office of Science. The SDSS-III web site is http://www.sdss3.org/.

SDSS-III is managed by the Astrophysical Research Consortium for the Participating Institutions of the SDSS-III Collaboration including the University of Arizona, the Brazilian Participation Group, Brookhaven National Laboratory, Carnegie Mellon University, University of Florida, the French Participation Group, the German Participation Group, Harvard University, the Instituto de Astrofisica de Canarias, the Michigan State/Notre Dame/JINA Participation Group, Johns Hopkins University, Lawrence Berkeley National Laboratory, Max Planck Institute for Astrophysics, Max Planck Institute for Extraterrestrial Physics, New Mexico State University, New York University, Ohio State University, Pennsylvania State University, University of Portsmouth, Princeton University, the Spanish Participation Group, University of Tokyo, University of Utah, Vanderbilt University, University of Virginia, University of Washington, and Yale University.

The Hobby-Eberly Telescope (HET) is a joint project of the University of Texas at Austin, the Pennsylvania State University, Stanford University, Ludwig-Maximilians-Universit\"at M\"unchen, and Georg-August-Universit\"at G\"ottingen. The HET is named in honor of its principal benefactors, William P. Hobby and Robert E. Eberly. 

The CSS survey is funded by the National Aeronautics and Space Administration under Grant No. NNG05GF22G issued through the Science Mission Directorate Near-Earth Objects Observations Program. The CRTS survey is supported by the U.S.~National Science Foundation under
grants AST-0909182 and AST-1313422.

\appendix

\section{Notes on Individual Objects}

\subsection{J030000.57+004828.0}

There are three available DR7 observations for J030000; to compare these epochs we scaled the two most recent spectra to coincide with the earliest observation. We did not detect any significant variability with the 8$\sigma$ criterion between any two DR7 epochs so we combined the three spectra together via a weighted average to be compared to the MDM observation; no significant variability was detected. The time-scales between any two epochs ranged from $\sim$35 d to $\sim$6.0 yr. 

\citet{viv12} also conducted a variability study on this source, utilizing observations in 2007 and 2008 in addition to the DR7 observations. They did not detect significant BAL variability across their $\sim$4.2 yr coverage, consistent with our results, however they report a possible variation of the feature between $\sim$2400--2500 \AA. They see a consistent rise in flux across this interval between DR7 and their 2008 spectra, however we do not see any change in our comparison between DR7 and 2012 (see Fig. 1). It is unclear which ionic species contribute to the absorption in this region; \citet{viv12} mention the possibility of a higher velocity component of the Fe\thinspace \textsc{ii} (UV1) multiplet. The lack of variability that we observe between 2000 and 2012 suggests the object varied over the time-scale probed by \citet{viv12} then went back to its original state; follow-up observations may shed light on this behavior.
 
 \subsection{J031856.62-060037.7}
 
 We have three observations for J031856 spanning time-scales from $\sim$3.4--3.8 yr. The wavelength coverage for studying variability was from $\sim$1300--2030 \AA, a range that included the resonance transitions Si\thinspace \textsc{iv} $\lambda\lambda$1393, 1402, C\thinspace \textsc{iv} $\lambda\lambda$1548, 1550, Al\thinspace \textsc{ii} $\lambda$1670, and Al\thinspace \textsc{iii} $\lambda\lambda$1854, 1862. The Fe\thinspace \textsc{ii} and Mg\thinspace \textsc{ii} transitions are found in redder regions of the spectrum outside of our observational coverage. We do not detect any significant absorption line variability using the 8$\sigma$ criterion for these ions across the observed time-scales, consistent with the results of \citet{viv12} who also study this source over $\sim$3 yr between 2001 and 2009 and had additional wavelength coverage extending out to $\sim$2850 \AA.
 
 \subsection{J084044.41+363327.8}
 
 Our sample contains eight spectra for J084044, spanning time-scales from $\sim$10 d to $\sim$5.3 yr. No variability was detected between the HET epochs using the 8$\sigma$ criterion so we combined them into one weighted-averaged spectrum for comparisons (weighted averages were also employed for HET epochs of J121442 and J142703, due to lack of variability).
 
Possible absorption line changes were detected over our available time-scales in scattered regions between $\sim$2210--2800 \AA. The region at $\sim$2800 \AA \ cannot be conclusively attributed to the Mg\thinspace \textsc{ii} $\lambda\lambda$2796, 2803 BAL due to contamination with the Mg\thinspace \textsc{ii} BEL. In addition, based upon follow-up inspection, we conclude that the variability elsewhere is potentially due to systematic uncertainties in matching instrumental resolutions.

 \citet{viv12} conducted a variability analysis on J084044 from 2001 to 2009 ($\sim$4 yr), finding no significant absorption line changes. Our lack of conclusive absorption line variability on time-scales of $\sim$0.85-5.3 yr is consistent with their result.
 
 \subsection{J112526.12+002901.3}
 
 The time-scales between our three available epochs for J112526 range from $\sim$0.55--6.4 yr. The only significant variability was detected between the DR7 and DR12 comparison, a timescale of $\sim$5.8 yr, with variable regions ranging from $\sim$2770--2810 \AA. This could correspond to a weakening in the Mg\thinspace \textsc{ii} $\lambda\lambda$2796, 2803 BAL and/or strengthening of the Mg\thinspace \textsc{ii} BEL. \citet{hal13} came to the same conclusion after scaling the DR12 spectrum using a function expressed as a constant times a power law, different than the first order linear spline function we used to scale the DR12 epoch. 
 
 Inspection of Fig. 1 shows that there are tentative decreases in absorption line strength coming from  the Fe\thinspace \textsc{ii} (UV1,2,3,62,63) multiplets. Because these regions only meet the 6$\sigma$ criterion and due to the contamination of the Mg\thinspace \textsc{ii} BAL with the Mg\thinspace \textsc{ii} BEL, we conclude that J112526 is a tentative case and the origin of the variability is ambiguous (although see \S 4.4 for discussion). 
 
 This source is one of the few BAL quasars for which a portion of the BAL is red-shifted \citep{hal13}. The velocity range over which we see Mg\thinspace \textsc{ii} absorption is from approximately $-1000$ to $+1400$ km s$^{-1}$. \citet{hal13} also mentioned that the blue-shifted part of the Mg\thinspace \textsc{ii} BAL weakened more than the red-shifted part, an observation that we also detect by eye. Two likely scenarios to explain red-shifted absorption could be either a rotationally dominated outflow, radial infall of gas or a combination of both (see \citealt{hal13} for details). Observing a higher degree of variability in the blue-shifted part of the Mg\thinspace \textsc{ii} BAL as opposed to the red-shifted part might indicate that there are multiple outflow--infall systems contributing to the absorption. 

\renewcommand{\thefigure}{\Alph{figure}}

\setcounter{figure}{0}

\begin{figure*}
\begin{center}
\includegraphics[width=2.0\columnwidth,angle=0]{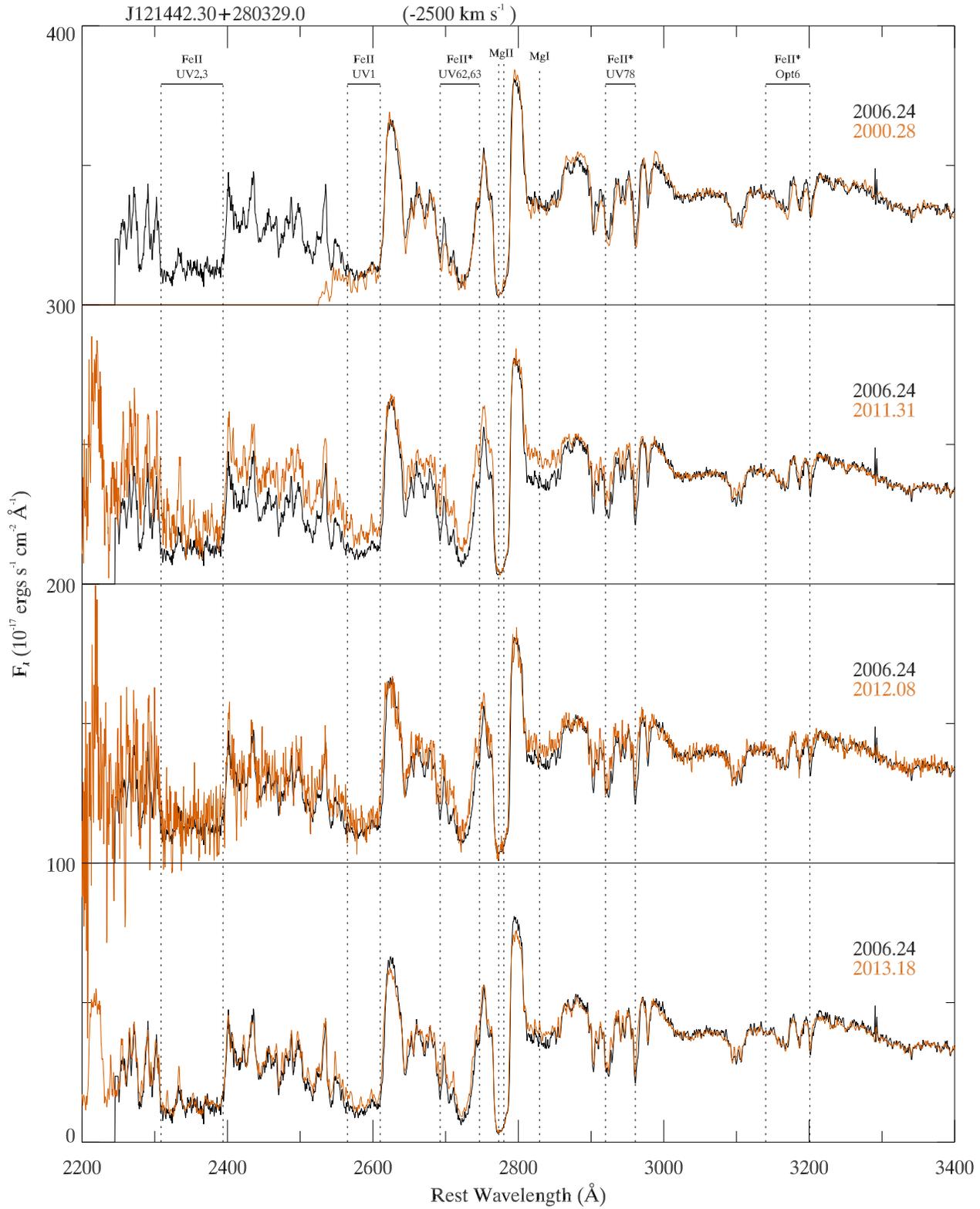}
\caption{Spectra comparisons of the FeLoBAL quasar J121442, revealing a change in the direction of absorption line variability. The top of the figure lists the object name and velocity shift of the overlaid line template (negative velocity indicates the lines are blue-shifted), which is shown in the topmost panel. Decimal years next to each comparison indicate the two epochs plotted, with the color of each year corresponding to the color of the associated spectrum. Solid horizontal lines represent zero points for each comparison.}
\end{center}
\end{figure*}

\begin{figure*}
\begin{center}
\includegraphics[width=2.0\columnwidth,angle=0]{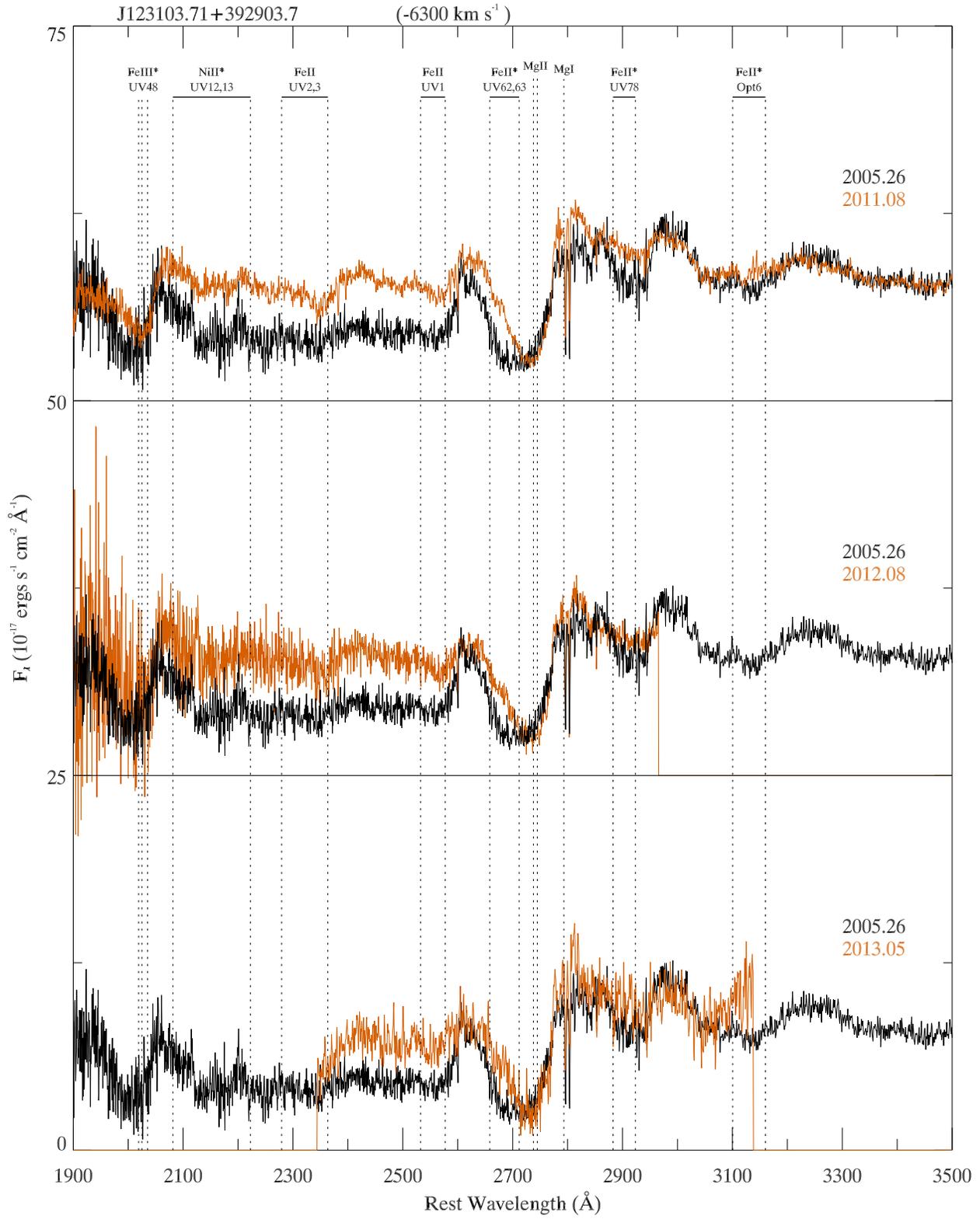}
\caption{Spectra comparisons of the FeLoBAL quasar J123103. The top of the figure lists the object name and velocity shift of the overlaid line template (negative velocity indicates the lines are blue-shifted), which is shown in the topmost panel. Decimal years next to each comparison indicate the two epochs plotted, with the color of each year corresponding to the color of the associated spectrum. Solid horizontal lines represent zero points for each comparison.}
\end{center}
\end{figure*}

\bsp

\label{lastpage}

\end{document}